\documentclass[reprint,scriptaddress,osajnl2,showkeys,showpacs]{revtex4-1}

\usepackage[latin1]{inputenc}
\usepackage{amsmath} 
\usepackage{amsfonts}
\usepackage{amssymb}
\usepackage{graphicx} 
\usepackage{lmodern} 
\usepackage{microtype} 
\usepackage[hyperindex,pdfborder={0 0 0}]{hyperref} 
\usepackage[draft,danish]{fixme} 
\usepackage{pgf,tikz} 
\usetikzlibrary{arrows} 

\newcommand{\Eqref}[1]{Eq.~\eqref{#1}} 
\newcommand{\Eqsref}[1]{Eqs.~\eqref{#1}} 
\newcommand{\EqsrefM}[2]{Eqs.~\eqref{#1}-\eqref{#2}} 
\newcommand{\Figref}[1]{Fig.~\ref{#1}} 
\newcommand{\enh}[1]{\text{ #1}}
\newcommand{\vekn}[1]{\boldsymbol{\mathrm{#1}}}  	
\newcommand{\dd}{\mathrm{d}}						
\newcommand{\ddd}{\:\mathrm{d}}						
\newcommand{\abs}[1]{\ensuremath{\left|#1 \right|}}	
\newcommand{\IP}[2]{\langle #1 | #2 \rangle}	
\newcommand{\vekr}{\vekn{r}} 
\newcommand{\vekR}{\vekn{R}} 
\newcommand{\vekE}{\vekn{E}} 
\newcommand{\vekEB}{\vekn{E}_{\mathrm{B}}} 
\newcommand{\deps}{\Delta \epsilon} 
\newcommand{\GB}{\vekn{G}_{\mathrm{B}}} 
\newcommand{\gB}{g_{\mathrm{B}}} 
\newcommand{\Lso}{\vekn{L}} 
\newcommand{\epsind}[1]{\epsilon_{\mathrm{\scriptscriptstyle{#1}}}}
\newcommand{\epsb}{\epsind{B}} 
\newcommand{\kb}{k_{\mathrm{B}}} 
\newcommand{\EpsL}{\mathcal{E}_{\mathrm{L}}} 
\newcommand{\EpsG}{\mathcal{E}_{\mathrm{G}}} 
\newcommand{\vekkb}{\vekk_{\mathrm{B}}}
\newcommand{\vekf}{\vekn{f}} 
\newcommand{\vekSB}{\vekn{S}_{\mathrm{B}}} 
\newcommand{\Cext}{C_{\mathrm{ext}}} 
\newcommand{\Qext}{Q_{\mathrm{ext}}} 
\newcommand{\omegat}{\tilde{\omega}}
\newcommand{\quasi}{\vekn{f}}

\newcommand{\eps}[1]{\epsilon_{#1}} 
\newcommand{\Cen}[1]{\vekn{r}_{#1}^0} 
\newcommand{\Rad}[1]{R_{#1}} 
\newcommand{\psij}[3]{\psi_{#2,#3}^{#1}} 
\newcommand{\psijNN}[3]{\tilde{\psi}_{#2,#3}^{#1}} 
\newcommand{\vekrj}[1]{\vekr_{#1}} 
\newcommand{\Nj}[2]{N_{#1}^{#2}} 
\newcommand{\jj}[1]{j_{#1}} 
\newcommand{\kj}[1]{k_{#1}} 
\newcommand{\rj}[1]{r_{#1}} 
\newcommand{\Thj}[1]{\theta_{#1}} 
\newcommand{\Phj}[1]{\phi_{#1}} 
\newcommand{\CC}[1]{\lbrace #1 \rbrace^{Y}} 
\newcommand{\Yj}[2]{Y_{#1}^{#2}} 
\newcommand{\psijb}[3]{\psi_{#2,#3}^{#1,\mathrm{B}}} 
\newcommand{\psijbNN}[3]{\tilde{\psi}_{#2,#3}^{#1,\mathrm{B}}} 
\newcommand{\Njb}[2]{N_{#1}^{#2, \mathrm{B}}} 
\newcommand{\phib}[2]{\varphi_{#1,#2}^{\mathrm{B}}} 
\newcommand{\hj}[1]{h_{#1}^{(1)}} 
\newcommand{\Dsum}[2]{\sum_{#1,#2}} 
\newcommand{\intSA}{\int_{\Omega}} 
\newcommand{\vekrjp}[1]{\vekr'_{#1}} 
\newcommand{\veke}[1]{\,\vekn{e}_{#1}} 
\newcommand{\jp}{j'} 
\newcommand{\alphap}{\alpha'} 
\newcommand{\lp}{l'} 
\newcommand{\mpr}{m'} 
\newcommand{\depsm}{\vekn{\deps}} 
\newcommand{\MB}{\vekn{M}_{\mathrm{B}}} 
\newcommand{\Gm}{\vekn{G}} 
\newcommand{\vekb}{\vekn{b}} 
\newcommand{\Smat}[4]{S_{#1, #3}^{#2, #4}} 
\newcommand{\Smath}[4]{\hat{S}_{#1, #3}^{#2, #4}} 
\newcommand{\Gaunt}[5]{\mathcal{G}(#1, #2; #3, #4; #5)} 
\newcommand{\Omeg}[2]{\Omega_{#1}^{#2}} 
\newcommand{\IA}[2]{I_{#2}^{\mathbb{R}^3-\delta V_{#1}}} 
\newcommand{\IB}[2]{I_{#2}^{\mathbb{R}^3-V_{#1}}} 
\newcommand{\kd}[2]{\delta_{#1 #2}} 
\newcommand{\vekk}{\vekn{k}} 
\newcommand{\vekeb}{\vekn{e}_{\mathrm{B}}} 
\newcommand{\vekrd}{\vekr_{\mathrm{d}}} 
\newcommand{\lmax}{l_{\mathrm{max}}} 

\newcommand{\ci}{\mathrm{i}} 
\newcommand{\nm}{\enh{nm}} 

\makeindex
\begin{document}

\title{Three-dimensional integral equation approach to light scattering, extinction cross sections, local density of states and quasinormal modes}
\author{Jakob Rosenkrantz de Lasson}
\email{jakob@jakobrdl.dk}
\author{Jesper M\o rk}
\author{Philip Tr\o st Kristensen}
\affiliation{DTU Fotonik, Department of Photonics Engineering, Technical University of Denmark, \O rsteds Plads, Building 343, DK-2800 Kongens Lyngby, Denmark}
\date{\today}


\begin{abstract}
We present a numerical formalism for solving the Lippmann-Schwinger equation for the electric field in three dimensions. The formalism may be applied to scatterers of different shapes and embedded in different background media, and we develop it in detail for the specific case of spherical scatterers in a homogeneous background medium. In addition, we show how several physically important quantities may readily be calculated with the formalism. These quantities include the extinction cross section, the total Green's tensor, the projected local density of states and the Purcell factor as well as the quasinormal modes of leaky resonators with the associated resonance frequencies and quality factors. We demonstrate the calculations for the well-known plasmonic dimer consisting of two silver nanoparticles and thus illustrate the versatility of the formalism for use in modeling of advanced nanophotonic devices.
\end{abstract}



\maketitle

\section{Introduction}
Realization of optical devices based on optical micro- or nanostructures such as photonic crystals~\cite{Bykov_SovJQuantum_4_861_1975, Yablonowitch_PRL_58_2059_1987,John_PRL_58_2486_1987} or plasmonic
nanoparticles~\cite{Catchpole,PlasmonRuler3D,Zhang2008} rely on a prolific interplay
between advanced fabrication techniques and accurate numerical methods. The latter paves the way for design of advanced optical functionalities as well as systematic
studies and in-depth understanding of the physical mechanisms at play. Additionally,
numerical modeling serves as an indispensable tool in the interpretation of
experimental results, and the study and development of numerical modeling methods therefore remain an important and integral part of modern nanophotonics research.  Propagation of light, in the form of electromagnetic fields, is governed by Maxwell's equations, and in spite of being known for more than a century these equations remain very difficult to solve and display rich behavior. Analytical solutions are available
only for a limited number of geometries, and numerical solvers are thus
indispensable in the design of practical devices. Each numerical scheme has advantages and limitations, as analyzed, for example, with Photonic-Crystal-based Vertical-Cavity
Surface-Emitting Lasers (PC-VCSELs) as benchmark structures using four different
methods in~\cite{Dems2010}. The most prominent advantages of the integral equation
approach that we present in this paper are versatility in the form of easy access to
figures of merit and high accuracy with a built-in error measure.

The most popular numerical methods in the field of nanophotonics are the finite-difference time-domain (FDTD) method~\cite{FDTD} and the finite element method (FEM)~\cite{FEM}, which are both based on spatial discretization of Maxwell's equations. FDTD uses a rectangular grid and a simple time-stepping procedure to evolve the fields in time, wheras FEM uses a non-uniform triangular meshing, which can more easily adapt to curved surfaces, and is most often used for frequency domain problems. As a powerful hybrid approach, discontinuous Galerkin methods use a variant of FEM based on non-overlapping basis functions leading to improved performance in time-domain calculations~\cite{Busch_LaserPhotRev_5_773_2011}. These methods can easily be adapted to treat arbitrary structures, but the necessity to discretize the entire space may in practice lead to large requirements in terms of memory and computational power, in particular for three-dimensional (3D) problems. Alternatives include modal expansion techniques such as the Fourier modal method~\cite{Moharam1995} and the Rayleigh multipole method~\cite{RayleighMultipole}, in which the fields are expanded on a chosen set of basis functions, and the electromagnetic boundary conditions (BCs) are satisfied to determine the expansion coefficients. In both FDTD, FEM and modal expansion techniques the need to minimize parasitic reflections from the calculation domain boundaries usually entails the introduction of perfectly matched layers (PMLs)~\cite{Berenger}. Another class of methods is based on surface or volume integrals~\cite{NovotnyFull}. One advantage of this procedure is that only bounded parts of space need to be discretized which prompts faster computations. Typical approaches employ expansions of the fields on orthonormal sets of basis functions, and the integral equations are converted into systems of linear equations for the expansion coefficients. A popular choice of basis functions is the so-called pulse basis functions~\cite{CompMethods} that form the foundation of the discrete dipole approximation~\cite{DDA}. The simplicity of these piecewise-constant basis functions allows a simple treatment of arbitrary geometries, but their simplicity in turn leads to very large systems of equations. More severely, the pulse basis functions induce fictitious current densities which lead to inaccurate results for high index contrasts~\cite{CompMethods}. 

In this work, we present a volume integral formulation based on the Lippmann-Schwinger equation~\cite{Levine1948} and the electromagnetic Green's tensor for the electric field. In this approach, the electric field satisfies the radiation condition~\cite{Jackson} by construction, and artificial BCs like PMLs are not needed. Known results for the Green's tensor may be used to model inhomogeneities embedded in different background environments such as homogeneous space or layered media~\cite{Tai_1994,NovotnyFull,LayeredGreens}. As an alternative to the pulse basis functions, we employ expansions in scalar wavefunctions that are solutions of a homogeneous Helmholtz equation~\cite{MultiScat}. The method was developed for 2D structures in~\cite{Philip2}; in this work we generalize the method to 3D and elaborate on the versatility of the method for calculating various physically important quantities. The general procedure can be applied to scatterers of different shapes in different background environments, but we focus here on the specific example of spherical particles embedded in a homogeneous background medium. In this case the calculation of the ensuing matrix elements dramatically simplifies and can be expressed analytically. The special case of scattering by a single particle is described by Mie scattering theory~\cite{Mie}, and generalized Mie theories for several spherical particles have been developed~\cite{MieGeneralized,TMatrix,Abajo}. Common ingredients in these schemes are expansions in scalar (or spherical) wavefunctions as well as the use of addition and translation theorems. These functions and theorems are also employed in the present formalism, but whereas the generalized Mie theories rely on explicit fulfillment of the electromagnetic BCs to determine expansion coefficients, our approach exploits the Lippmann-Schwinger equation that implicitly satisfies the BCs. The present formalism is advantageous as it gives direct access to a large number of physically relevant quantities, such as the electric field --- including the near- and far-fields~\cite{NovotnyFull} --- the extinction, scattering and absorption cross sections~\cite{BornWolf}, the total Green's tensor for background medium plus scatterers~\cite{Tai_1994,Martin1}, the projected local density of states~\cite{Sprik_EuroPhys_35_265_1996,NovotnyFull}, the Purcell factor~\cite{Purcell} as well as quasinormal modes (or cavity modes) with their associated $Q$-factors~\cite{Lee_JOSAB_16_1409_1999,ModeVolume,NovotnyFull}. Additionally, the formalism contains an explicit error estimate that we demonstrate and discuss. As an application of the formalism, we consider a plasmonic dimer. This system has been widely studied, both experimentally and theoretically, and we choose this well-known material configuration to display how the different physically important quantities can be directly analyzed with the method. Calculation examples using up to 20 particles are demonstrated in~\cite{deLasson2012}. 

The article is organized as follows: Section~\ref{Sec:MutiScatFormalism} presents the details of the formalism, including the introduction of the Lippmann-Schwinger equation, the expansion of the electric field and the evaluation of the matrix elements. It is shown how the extinction efficiency, the Green's tensor and the projected local density of states are obtained within our framework. Section~\ref{Sec:PlasmonicDimer} provides example calculations for a plasmonic dimer. Specifically, plane wave scattering on the dimer is demonstrated via the excitation of localized surface plasmons in the vicinity of the dimer, and extinction efficiency spectra are presented and discussed. Furthermore, calculations of the Green's tensor and the Purcell factor for the dimer are performed, and two different modes of the dimer are determined and visualized. Finally, Section~\ref{Sec:Conclusion} concludes the work. A number of appendices give various definitions and detailed expressions, and it is our hope that the interested reader will be able to implement and apply the formalism with a relatively small effort.

\section{Multiple-Scattering Formalism} \label{Sec:MutiScatFormalism}
\subsection{Lippmann-Schwinger Equation} \label{Sec:ProblemFormulation}
\begin{figure}[htb!]
\centering
\hspace{-2cm}
\begin{tikzpicture}[line cap=round,line join=round,>=triangle 45,x=1.0cm,y=1.0cm, transform canvas={scale=0.62},font=\LARGE,line width=0.25mm]
\newcommand\dxg{1}
\newcommand\dyg{0}
\newcommand\dxj{-3}
\newcommand\dyj{-1}
\newcommand\dxjp{6}
\newcommand\dyjp{0}
\newcommand\dxjt{2}
\newcommand\dyjt{-4}

\shade[ball color=blue!10!white,opacity=0.20] (\dxj+0,\dyj+0) circle (1cm);
\shade[ball color=blue!10!white,opacity=0.20] (\dxjp+0,\dyjp+0) circle (1cm);
\shade[ball color=blue!10!white,opacity=0.20] (\dxjt+0,\dyjt+0) circle (1cm);

\draw [->,thick,color=blue,-latex] (\dxj+0,\dyj+0)  -- (\dxj+0,\dyj+2) node(yline)[right]{$z_1$}; 
\draw [->,thick,color=blue,-latex] (\dxj+0,\dyj+0)  -- (\dxj+1.8,\dyj-0.7) node(xline)[right]{$y_1$};
\draw [->,thick,color=blue,-latex] (\dxj+0,\dyj+0)  -- (\dxj-1.8,\dyj-0.7) node(xline)[below]{$x_3$}; 
\fill [color=black] (\dxj+0,\dyj+0) circle (2.5pt); 
\draw (\dxj-1,\dyj+0) arc (180:360:1cm and 0.5cm);
\draw[dashed] (\dxj-1,\dyj+0) arc (180:0:1cm and 0.5cm);
\draw[dashed] (\dxj+0,\dyj+1) arc (90:-90:0.5cm and 1cm);
\draw (\dxj+0,\dyj+1) arc (90:270:0.5cm and 1cm);
\draw (\dxj+0,\dyj+0) circle (1cm);

\draw [->,thick,color=blue,-latex] (\dxjp+0,\dyjp+0)  -- (\dxjp+0,\dyjp+2) node(yline)[right]{$z_{2}$}; 
\draw [->,thick,color=blue,-latex] (\dxjp+0,\dyjp+0)  -- (\dxjp+1.8,\dyjp-0.7) node(xline)[right]{$y_{2}$};
\draw [->,thick,color=blue,-latex] (\dxjp+0,\dyjp+0)  -- (\dxjp-1.8,\dyjp-0.7) node(xline)[below]{$x_{2}$};
\fill [color=black] (\dxjp+0,\dyjp+0) circle (2.5pt); 
\draw (\dxjp-1,\dyjp+0) arc (180:360:1cm and 0.5cm);
\draw[dashed] (\dxjp-1,\dyjp+0) arc (180:0:1cm and 0.5cm);
\draw[dashed] (\dxjp+0,\dyjp+1) arc (90:-90:0.5cm and 1cm);
\draw (\dxjp+0,\dyjp+1) arc (90:270:0.5cm and 1cm);
\draw (\dxjp+0,\dyjp+0) circle (1cm);

\draw [->,thick,color=blue,-latex] (\dxjt+0,\dyjt+0)  -- (\dxjt+0,\dyjt+2) node(yline)[right]{$z_{3}$}; 
\draw [->,thick,color=blue,-latex] (\dxjt+0,\dyjt+0)  -- (\dxjt+1.8,\dyjt-0.7) node(xline)[right]{$y_{3}$};
\draw [->,thick,color=blue,-latex] (\dxjt+0,\dyjt+0)  -- (\dxjt-1.8,\dyjt-0.7) node(xline)[below]{$x_{3}$};
\fill [color=black] (\dxjt+0,\dyjt+0) circle (2.5pt); 
\draw (\dxjt-1,\dyjt+0) arc (180:360:1cm and 0.5cm);
\draw[dashed] (\dxjt-1,\dyjt+0) arc (180:0:1cm and 0.5cm);
\draw[dashed] (\dxjt+0,\dyjt+1) arc (90:-90:0.5cm and 1cm);
\draw (\dxjt+0,\dyjt+1) arc (90:270:0.5cm and 1cm);
\draw (\dxjt+0,\dyjt+0) circle (1cm);

\draw (0+\dxg,-0.80+\dyg) node[circle]{$\epsb$}; 
\draw (0+\dxj,-0.72+\dyj) node[circle]{$\eps{1}$}; 
\draw (0+\dxjp,-0.72+\dyjp) node[circle]{$\eps{2}$}; 
\draw (0+\dxjt,-0.72+\dyjt) node[circle]{$\eps{3}$}; 

\draw[domain=0:3.14,samples=50,color=black]	plot [id=sin,xshift=1cm,yshift=1cm,rotate=135] (\x,{0.8*sin(\x r*5)}) node[right] {};
\draw[->,color=black,line width=1pt] (1.15,0.85) -- (1.8,0.2);
\draw (0.7,2.8) node[circle]{\rotatebox{-45}{$\vekE_{\mathrm{B}}$}};

\end{tikzpicture}
\vspace{3.5cm}
\caption{(Color online) Example of scattering geometry where an incoming field, $\vekEB$, impinges on $N = 3$ spherical scatterers embedded in a homogeneous material of permittivity $\epsb$. The scatterers have permittivities $\eps{j}$, and the local coordinate systems are indicated.} \label{Fig:Spheres3DGeneric} 
\end{figure}
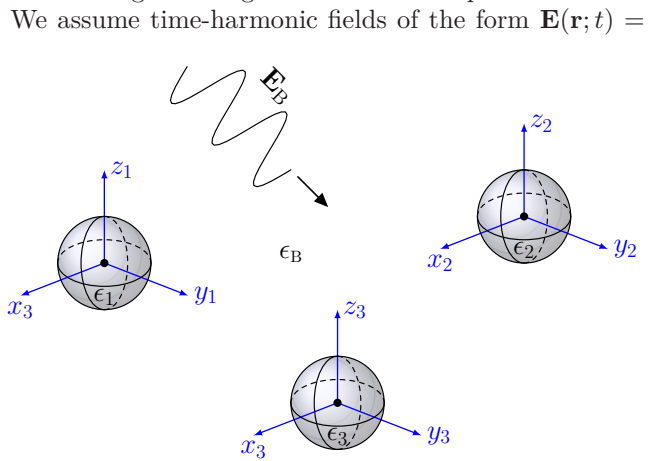

We consider scattering of an incoming electric field, $\vekEB$, on $N$ spherical scattering objects embedded in a homogeneous space of relative permittivity $\epsb(\omega)$. The scattering objects have relative permittivities $\eps{j}(\omega), j = 1,2, \dots, N$, and we assume non-magnetic, isotropic scatterers throughout. \Figref{Fig:Spheres3DGeneric} shows an example with $N = 3$.
\vspace{2cm}
We assume time-harmonic fields of the form $\vekE(\vekr;t) = \vekE(\vekr;\omega) \exp(-\ci\omega t)$, where $\omega$ is the angular frequency, and where the fields $\vekE(\vekr;\omega)$ solve the wave equation
\begin{align}\label{Eqn:WaveEqn}
\nabla \times \nabla \times \vekE(\vekr;\omega) - k_0^2 \epsb(\omega) \vekE(\vekr;\omega) = k_0^2 
\deps(\vekr;\omega) \vekE(\vekr;\omega),
\end{align}
in which $k_0 = \omega/c$, $c$ being the speed of light in free-space, and $\deps(\vekr;\omega) \equiv \epsilon(\vekr;\omega) - \epsb(\omega)$ are the 
free-space wave number and the relative permittivity contrast, respectively. We suppress the explicit frequency dependence onwards. The solution of \Eqref{Eqn:WaveEqn} is the sum of the incoming field and the scattered field, giving rise to the Lippmann-Schwinger equation~\cite{Levine1948,Martin1}
\begin{align}\label{Eqn:LSEqn}
\vekE(\vekr) = \vekEB(\vekr) + k_0^2 \int_{V_{\mathrm{scat}}} \GB(\vekr,\vekr') 
\deps(\vekr') \vekE(\vekr') \ddd \vekr',
\end{align}
where $\GB(\vekr,\vekr')$ and $V_{\mathrm{scat}}$ are the electromagnetic Green's tensor of the background medium and the volume of the scattering objects, respectively. The former satisfies \Eqref{Eqn:WaveEqn} with the right hand side replaced by $\vekn{I} \delta(\vekr-\vekr')$, $\vekn{I}$ being a unit dyad. The elements of the homogeneous background Green's tensor can be expressed as~\cite{NovotnyFull}
\begin{subequations} \label{Eqn:GreensDefs}
\begin{align}
\GB^{\alpha \alphap}(\vekr,\vekr') &= \left(\kd{\alpha}{\alphap} + \frac{1}{\kb^2} \partial_{\alpha} \partial_{\alphap} \right) \gB(\vekr, \vekr'), \label{Eqn:GreensTensor} \\
\gB(\vekr, \vekr') &=\frac{\exp\left(\ci \kb \abs{\vekr - \vekr'} \right)}{4\pi \abs{\vekr-\vekr'}}, \label{Eqn:GreensFunction}
\end{align}
\end{subequations}
where $\gB(\vekr, \vekr')$ and $\kb \equiv \sqrt{\epsb} k_0$ are the scalar Green's function and the wave number of the background medium, respectively, and where $\partial_{\alpha} \equiv \partial / \partial \alpha$ with $\alpha, \alphap \in \lbrace x, y, z \rbrace$, and $\kd{\alpha}{\alpha'}$ denotes the Kronecker delta. 
$\GB(\vekr,\vekr')$ diverges at $\vekr = \vekr'$, and since we are interested in determining the field inside the scattering objects, this singularity must be isolated. This is done using the following altered Lippmann-Schwinger equation~\cite{Yaghjian}
\begin{align} \label{Eqn:LSEqnPrin}
\vekE(\vekr) = \vekEB(\vekr) + k_0^2 \int_{V_{\mathrm{scat}}-\delta V} 
&\GB(\vekr,\vekr') \deps(\vekr') \vekE(\vekr') \ddd \vekr' \nonumber \\
&-\Lso \frac{\deps(\vekr)}{\epsb}\vekE(\vekr).
\end{align}
The integral is now evaluated as a principal value, omitting the point $\vekr = \vekr'$, which is compensated by the introduction of the source dyadic $\Lso$. We choose spherical exclusion volumes $\delta V$, for which $\Lso^{\alpha \alpha'} = \kd{\alpha}{\alpha'}/3$~\cite{Yaghjian}. We note that \Eqref{Eqn:LSEqnPrin} is implicit for $\vekr \in V_{\mathrm{scat}}$ and explicit for $\vekr \notin V_{\mathrm{scat}}$. Therefore, the majority of the computational work will concern the calculation of the field inside the scatterers. Once the field is known in these regions, it is straightforward to calculate the solution at all other points. In Sections~\ref{Sec:ExpansionField} through \ref{Sec:BackgroundField}, we present the procedure for solving \Eqref{Eqn:LSEqnPrin}; in Section~\ref{Sec:SummaryFormalism} the steps in the procedure are summarized.

If the field can be well approximated by constants inside each of the scatterers, \Eqref{Eqn:LSEqnPrin} may be solved simply by pulling the field outside the integral. This transforms the implicit Lippmann-Schwinger equation into a system of algebraic equations for the $3N$ field values inside the scatterers, which may be solved directly. We refer to this approximate solution scheme as the dipole approximation (DA), and in Section~\ref{Sec:PlaneWaveScattering} we compare it to the full formalism that we develop below. Note that we take the finite extent of the scatterers into account by evaluating the integrals over the elements of the Green's tensor analytically in the dipole approximation presented here. This approach is therefore more elaborate than the well-known point-scatterer model~\cite{MetaBook} that is often employed in the literature. An extensive work on point-scatterer modeling can be found in~\cite{Shore2007}.

\subsection{Expansion of Electric Field} \label{Sec:ExpansionField}
To solve \Eqref{Eqn:LSEqnPrin} inside $V_{\mathrm{scat}}$, we employ an expansion of the field and the background field inside scatterer $j$,
\begin{subequations} \label{Eqn:FieldExpansions}
\begin{align}
\vekE(\vekrj{j}) &= \sum_{\alpha,l,m} a_{j \alpha l m} \psij{j}{l}{m}(\vekrj{j}) \veke{\alpha}, \label{Eqn:FieldExpan} \\
\vekEB(\vekrj{j}) &= \sum_{\alpha,l,m} a_{j \alpha l m}^{\mathrm{B}} \psijb{j}{l}{m}(\vekrj{j}) \veke{\alpha}, \label{Eqn:FieldBExpan}
\end{align}
\end{subequations}
where $\Dsum{l}{m} \equiv \sum_{l=0}^{\infty} \sum_{m=-l}^{l}$, and $\veke{\alpha}$ is a unit polarization vector of the Cartesian direction $\alpha$. The basis functions $\psij{j}{l}{m}(\vekrj{j})$ and $\psijb{j}{l}{m}(\vekrj{j})$ are so-called spherical wavefunctions defined in the local coordinate system of the $j$th scatterer (coordinate axes in \Figref{Fig:Spheres3DGeneric}). These functions are defined in Appendix~\ref{App:SphWaveFunctions} and each include a spherical harmonic, $\Yj{l}{m}(\Thj{j},\Phj{j})$. The parameters $a_{j \alpha l m}$ ($a_{j \alpha l m}^{\mathrm{B}}$) are unknown (known) expansion coefficients, and the analysis is concerned with finding all $a_{j \alpha l m}$. 
We define an inner product
\begin{align}
\IP{f}{g} \equiv \int \CC{f(\vekr)} g(\vekr) \ddd \vekr,
\end{align}
where $\CC{f(\vekr)}$ implies that $f(\vekr)$ is complex conjugated in the spherical harmonic only. We then have
\begin{subequations} \label{Eqn:IP}
\begin{align}
\IP{\psij{j}{l}{m}}{\psij{\jp}{\lp}{\mpr}} &= \delta_{j j'} \delta_{l l'} \delta_{m m'}, \label{Eqn:BasisIP} \\
\IP{\psijb{j}{l}{m}}{\psijb{\jp}{\lp}{\mpr}} &= \delta_{j j'} \delta_{l l'} \delta_{m m'}, \label{Eqn:BackBasisIP} \\
\IP{\psij{j}{l}{m}}{\psijb{\jp}{\lp}{\mpr}} &= M_l^j \delta_{j j'} \delta_{l l'} \delta_{m m'}, \label{Eqn:BasisBackIP}
\end{align}
\end{subequations}
where $M_l^j$ is the overlap integral of two spherical Bessel functions.
Inserting the expansions in \Eqsref{Eqn:FieldExpansions} into \Eqref{Eqn:LSEqnPrin}, projecting onto $\psij{j}{l}{m}(\vekrj{j}) \veke{\alpha}$ and summing over all free indices produces the matrix equation
\begin{align} \label{Eqn:LSVecEqn}
\vekn{a} = \MB \vekn{a}_{\mathrm{B}} + \left(k_0^2 \Gm \depsm -  \frac{L}{\epsb} \depsm\right) \vekn{a},
\end{align}
where $\vekn{a}$ and $\vekn{a}_{\mathrm{B}}$ contain the expansion coefficients of the field and the background field, respectively. $L=1/3$ is a diagonal element of $\Lso$, while $\MB$ and $\depsm$ are diagonal matrices with diagonal elements $M_l^j$ and $\deps_j \equiv \eps{j}-\epsb$, respectively. Finally, $\Gm$ is a matrix with elements of the form
\begin{align} \label{Eqn:MatrixElement}
\Big [\Gm_{j,\jp}^{\alpha \alphap} \Big ]_{l,l'}^{m,m'} \equiv \int_{V_j} & \int_{V_{\jp} - \delta V} 
\CC{\psij{j}{l}{m}(\vekrj{j})} \GB^{\alpha \alphap}(\vekrj{j},\vekrjp{\jp}) \nonumber \\ 
&\times \psij{\jp}{\lp}{\mpr}(\vekrj{\jp}) \ddd \vekrjp{\jp} \ddd \vekrj{j}.
\end{align}
The expression for the matrix element in \Eqref{Eqn:MatrixElement} is independent of the shape of the scattering objects. As we demonstrate in the following section, the integrals can be solved analytically for spherical scatterers, but arbitrarily shaped scatterers can in principle be handled by evaluation of the integrals in \Eqref{Eqn:MatrixElement}. This is an advantage of the use of a volume integral formulation as compared, for instance, to the generalized Mie scattering theories that rely explicitly on the spherical shape of the scatterers.
\subsection{Green's Tensor Matrix Elements}\label{Sec:GreenMatElem}
The matrix elements in \Eqref{Eqn:MatrixElement} fall in two classes: When $j \neq \jp$, one has $\vekr \neq \vekr'$ by construction, and there is no need for a principal volume. Conversely, when $j = \jp$ we must treat the principal volume integral with care. We term these two classes of matrix elements scattering terms and self terms, respectively.
\subsubsection{Scattering Terms} \label{Sec:ScatTerms}
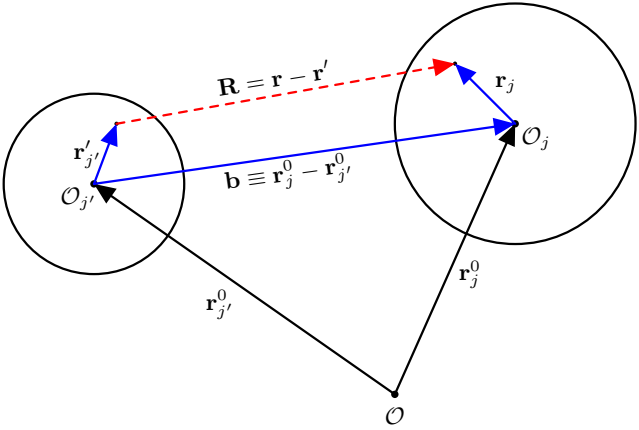
\begin{figure}[htb!]
\centering
\begin{tikzpicture} [line cap=round,line join=round,>=triangle 45,x=1.0cm,y=1.0cm,scale=0.4,line width=0.3mm]
\draw (14,7) circle (4cm);
\draw (0,5) circle (3cm);

\fill [color=black] (10,-2) circle (3.5pt);
\draw (10,-2.7) node[circle]{$\mathcal{O}$};
\fill [color=black] (14,7) circle (3.5pt);
\draw (14.7,6.5) node[circle]{$\mathcal{O}_j$};
\fill [color=black] (0,5) circle (3.5pt);
\draw (-0.5,4.5) node[circle]{$\mathcal{O}_{\jp}$};

\fill [color=black] (0.75,7) circle (2pt);
\fill [color=black] (12,9) circle (2pt);

\draw [->] (10,-2) -- (14,7); 
\draw [->] (10,-2) -- (0,5);
\draw [->,color=blue] (0,5) -- (14,7);
\draw [->,color=blue] (0,5) -- (0.75,7);
\draw [->,color=blue] (14,7) -- (12,9);
\draw [->,color=red,dashed] (0.75,7) -- (12,9);

\draw (13.7,8.3) node[circle]{$\vekrj{j}$};
\draw (-0.2,6) node[circle]{$\vekrjp{\jp}$};
\draw (12.5,2) node[circle]{$\Cen{j}$};
\draw (4.2,1) node[circle]{$\Cen{\jp}$};
\draw (6,8.5) node[circle]{\rotatebox{9}{$\vekR = \vekr - \vekr'$}};
\draw (6.5,5.3) node[circle]{\rotatebox{8}{$\vekn{b} \equiv \Cen{j} - \Cen{\jp}$}};

\end{tikzpicture}
\caption{(Color online) Two scatterers with indices $j$ and $\jp$, sketched in 2D. The centers of the scatterers, $\Cen{j}$ and $\Cen{\jp}$,  and two arbitrary local points inside the scatterers, $\vekrj{j}$ and $\vekrjp{\jp}$, are indicated. The displacement vectors between the two points, $\vekR$, and between the centers of the scatterers, $\vekb$, are shown.} \label{Fig:Scatterersjjp} 
\end{figure}

Fig.~\ref{Fig:Scatterersjjp} illustrates two spherical scatterers with indices $j$ and $\jp$. The displacement vector between two arbitrary points inside the scatterers, $\vekR$ (dashed vector), may be expressed using the three full vectors as $\vekR = \vekrj{j} + \vekn{b} - \vekrjp{\jp}$. 
This allows an expansion of the scalar Green's function $\gB(\vekr, \vekr')$ using the two-center expansion~\cite{MultiScat}
\begin{align} \label{Eqn:gBTwoCenter}
\gB(\vekr, \vekr') = \ci \kb \Dsum{p}{t} \Dsum{\nu}{\mu} \Smat{p}{t}{\nu}{\mu}(\vekb) \CC{\psijbNN{\jp}{p}{t}(\vekrjp{\jp})} \psijbNN{j}{\nu}{\mu}(\vekrj{j}),
\end{align}
where $\psijbNN{j}{l}{m}(\vekrj{j}) \equiv \psijb{j}{l}{m}(\vekrj{j})/\Njb{l}{j}$. The normalization constants $\Njb{l}{j}$ and separation matrices  $\Smat{p}{t}{\nu}{\mu}(\vekb)$ are defined in Appendices~\ref{App:SphWaveFunctions} and \ref{App:SepMatrices}, respectively. In this way we have essentially represented $\gB(\vekr, \vekr')$ using the local background basis functions. The elements of the Green's tensor then follow from \Eqref{Eqn:GreensTensor} as
\begin{align} \label{Eqn:GreensTenScat2}
\GB^{\alpha \alphap}(\vekr,\vekr') = \ci \kb& \Dsum{p}{t} \Dsum{\nu}{\mu} \Smat{p}{t}{\nu}{\mu}(\vekb) \CC{\psijbNN{\jp}{p}{t}(\vekrjp{\jp})} \nonumber \\
&\times \left(\kd{\alpha}{\alphap} + \frac{1}{\kb^2} \partial_{\alpha} \partial_{\alphap} \right) \psijbNN{j}{\nu}{\mu}(\vekrj{j}).
\end{align}
The Cartesian partial derivatives of the spherical wavefunctions may be expressed as sums of other spherical wavefunctions. We express this symbolically as
\begin{align} \label{Eqn:PartPartSphWaveSymbol}
\partial_{\alpha} \partial_{\alphap} \psijbNN{j}{\nu}{\mu}(\vekrj{j}) = \sum_{\gamma_{\alpha,\alphap}} g_{\gamma_{\alpha,\alphap}} \psijbNN{j}{\nu(\gamma_{\alpha,\alphap})}{\mu(\gamma_{\alpha,\alphap})}(\vekrj{j}),
\end{align}
where $\nu(\gamma_{\alpha,\alphap}) \equiv \nu + \gamma'_{\alpha,\alphap}$ and $\mu(\gamma_{\alpha,\alphap}) \equiv \mu + \gamma''_{\alpha,\alphap}$, with $\gamma'_{\alpha,\alphap}$ and $\gamma''_{\alpha,\alphap}$ being integers. The sum contains a finite number of terms, and the coefficients $g_{\gamma_{\alpha,\alphap}}$ depend both on the two polarizations $\alpha$ and $\alphap$ and on the basis function indices $\nu$ and $\mu$. The latter dependence, however, is suppressed for brevity. The detailed expressions for $g_{\gamma_{\alpha,\alphap}}$ are discussed in Appendix~\ref{App:CarDerivSphWave}. Combining Eqs.~\eqref{Eqn:MatrixElement}, \eqref{Eqn:GreensTenScat2} and \eqref{Eqn:PartPartSphWaveSymbol} and exploiting the orthogonality in \Eqref{Eqn:BasisBackIP} we have an explicit expression for a generic scattering matrix element
\begin{align}
 \Big [\Gm_{j,\jp}^{\alpha \alphap} \Big ]_{l,l'}^{m,m'} &= \ci \kb M_{\lp}^{\jp} M_{l}^{j} \Big / \left(\Njb{\lp}{\jp} \Njb{l}{j} \right) \nonumber \\
 &\hspace{1cm} \times \bigg(\kd{\alpha}{\alphap} \Smat{\lp}{\mpr}{l}{m}(\vekb) + \frac{1}{\kb^2} \sum_{\gamma_{\alpha,\alphap}} g_{\gamma_{\alpha,\alphap}} \nonumber \\
&\hspace{1.5cm} \times \Smat{\lp}{\mpr}{l-\gamma'_{\alpha,\alphap}}{m-\gamma''_{\alpha,\alphap}}(\vekb)  \bigg).
\end{align}
The computation of the separation matrices involved in the scattering matrix elements may be rather time consuming, and in practice accounts for the majority of the computation time. These calculations, however, can be optimized~\cite{PMartinSepMat}.
\subsubsection{Self Terms} \label{Sec:SelfTerms}
\begin{figure}[htb!]
\centering
\begin{tikzpicture} [line cap=round,line join=round,>=triangle 45,x=1.0cm,y=1.0cm, scale = 0.32,line width=0.3mm]

\fill [color=black!20] (6,19-0.5) circle (0.2cm);
\draw [dashed, dash pattern=on 7pt off 7pt] (8,14) circle (6cm);
\draw [color=red] (6,19-0.5) circle (0.2cm);

%
\fill [color=black] (8,14) circle (3.5pt);
\draw (8.7,13.5) node[circle]{$\mathcal{O}_j$};

\fill [color=black] (6,19-0.5) circle (2pt);
\fill [color=black] (12.5,16.5-0.5) circle (2pt);

\draw [->,color=blue] (8,14) -- (6,19-0.5);
\draw [->,color=blue] (8,14) -- (12.5,16.5-0.5);
\draw [->,color=red,dashed] (6,19-0.5) -- (12.5,16.5-0.5);

\draw (6.3,16.3-0.5) node[circle]{$\vekrj{j}$};
\draw (11.2,14.8-0.5) node[circle]{$\vekrjp{j}$};
\draw (9.3,18.6-0.5) node[circle]{\rotatebox{-21}{$\vekR' = \vekrjp{j} - \vekrj{j}$}};
\draw (8,6) node[circle]{$\Big [A_{j,j}^{\alpha \alphap} \Big ]_{l,l'}^{m,m'}$};

\end{tikzpicture}
\hspace{0.25cm}
\begin{tikzpicture} [line cap=round,line join=round,>=triangle 45,x=1cm,y=1cm, scale = 0.32,line width=0.3mm]
\fill [color=black!20] (8,14) circle (6cm);
\draw (8,14) circle (6cm);
\draw [color=red] (6,19) circle (0.2cm);

%
\fill [color=black] (8,14) circle (3.5pt);
\draw (8.7,13.5) node[circle]{$\mathcal{O}_j$};

\fill [color=black] (6,19) circle (2pt);
\fill [color=black] (14,18.5) circle (2pt);

\draw [->,color=blue] (8,14) -- (6,19);
\draw [->,color=blue] (8,14) -- (14,18.5);
\draw [->,color=red,dashed] (6,19) -- (14,18.5);

\draw (6.3,16.3) node[circle]{$\vekrj{j}$};
\draw (14,17.5) node[circle]{$\vekrjp{j}$};
\draw (9.5,18) node[circle]{\rotatebox{-4}{$\vekR' = \vekrjp{j} - \vekrj{j}$}};
\draw (8,6) node[circle]{$\Big [B_{j,j}^{\alpha \alphap} \Big ]_{l,l'}^{m,m'}$};

\end{tikzpicture}
\caption{(Color online) Integration domains for evaluation of self terms. Gray shading indicates a volume that is excluded from the $\vekrjp{j}$-integration. Left panel: Integration domain for $A_{j,j}^{\alpha \alphap}$, extending over all space minus the principal volume. Right panel: Integration domain for $B_{j,j}^{\alpha \alphap}$, extending over all space minus the scatterer volume.}
\label{Fig:Scatterersjj} 
\end{figure}
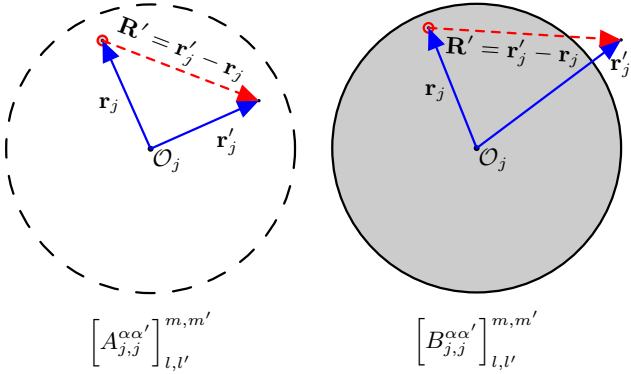

To evaluate the self terms where $j = \jp$, we split the $\vekrjp{j}$-integration into two parts, in analogy with the procedure in the 2D case in~\cite{Philip2}:
\begin{align}
\Big [\Gm_{j,j}^{\alpha \alphap} \Big ]_{l,l'}^{m,m'} = \Big [A_{j,j}^{\alpha \alphap} \Big ]_{l,l'}^{m,m'} - \Big [B_{j,j}^{\alpha \alphap} \Big ]_{l,l'}^{m,m'},
\end{align}
where
\begin{subequations}
\begin{align}
\Big [A_{j,j}^{\alpha \alphap} \Big ]_{l,l'}^{m,m'} &\equiv \int_{V_j} \int_{\mathbb{R}^3-\delta V_j} \CC{\psij{j}{l}{m}(\vekrj{j})} \GB^{\alpha \alphap}(\vekrj{j},\vekrjp{j}) \label{Eqn:Adef} \nonumber \\ 
&\hspace{1cm} \times \psij{j}{\lp}{\mpr}(\vekrjp{j}) \ddd \vekrjp{j} \ddd \vekrj{j}, \\
\Big [B_{j,j}^{\alpha \alphap} \Big ]_{l,l'}^{m,m'} &\equiv \int_{V_j} \int_{\mathbb{R}^3-V_j} \CC{\psij{j}{l}{m}(\vekrj{j})} \GB^{\alpha \alphap}(\vekrj{j},\vekrjp{j}) \label{Eqn:Bdef} \nonumber \\ 
&\hspace{1cm} \times \psij{j}{\lp}{\mpr}(\vekrjp{j}) \ddd \vekrjp{j} \ddd \vekrj{j}.
\end{align}
\end{subequations}
The integration domains for the two different cases are sketched in \Figref{Fig:Scatterersjj}. In both cases, gray shading indicates volumes that are excluded from the $\vekrjp{j}$-integrations.

The left panel in \Figref{Fig:Scatterersjj} illustrates the integration procedure for the elements  $A_{j,j}^{\alpha \alphap}$. For any fixed $\vekrj{j}$, we may equally well integrate $\vekR'$ or $\vekrjp{j}$ over the entire space minus the principal volume. To this end, we expand $\psij{j}{\lp}{\mpr}(\vekrjp{j})$ around $\vekrj{j}$ as~\cite{MultiScat}
\begin{align} \label{Eqn:ExpanBasisrj}
\psij{j}{\lp}{\mpr}(\vekrjp{j}) = \Nj{\lp}{j} \Dsum{\nu}{\mu} (-1)^{\mu} \Smath{\lp}{\mpr}{\nu}{\mu}(\vekrj{j}) \CC{\psijNN{j}{\nu}{-\mu}(\vekR')},
\end{align}
with the separation matrices $\Smath{\lp}{\mpr}{\nu}{\mu}(\vekrj{j})$ defined in Appendix~\ref{App:SepMatrices}. To express the elements of the Green's tensor we write the scalar Green's function as an outgoing spherical wavefunction (see definition in Appendix~\ref{App:SphWaveFunctions}) as
\begin{align}
\gB(\vekr,\vekr') = \gB(\vekR') = \frac{\ci \kb}{\sqrt{4 \pi}} \phib{0}{0}(\vekR').
\end{align}
The outgoing spherical wavefunctions satisfy a relation similar to \Eqref{Eqn:PartPartSphWaveSymbol}. Using this fact, and collecting from Eqs.~\eqref{Eqn:Adef} and \eqref{Eqn:ExpanBasisrj}, we may express the elements analytically as
\begin{align} \label{Eqn:AGeneric}
\Big [A_{j,j}^{\alpha \alphap} \Big ]_{l,l'}^{m,m'} &= \frac{\ci \kb}{\sqrt{4 \pi}} \Nj{\lp}{j} \bigg(\kd{\alpha}{\alphap} \IP{\psij{j}{l}{m}}{\Smath{\lp}{\mpr}{0}{0}} \IA{j}{0} \nonumber \\
&\hspace{0.5cm} + \frac{1}{\kb^2} \sum_{\gamma_{\alpha,\alphap}} g_{\gamma_{\alpha,\alphap}} (-1)^{\gamma''_{\alpha,\alphap}} \nonumber \\
&\hspace{0.5cm} \times \IP{\psij{j}{l}{m}}{\Smath{\lp}{\mpr}{\gamma'_{\alpha,\alphap}}{-\gamma''_{\alpha,\alphap}}} \IA{j}{\gamma'_{\alpha,\alphap}} \bigg ),
\end{align}
where the integral $\IA{j}{l}$ is defined and expressed in Appendix~\ref{App:RadIntegrals}. 

For the elements $B_{j,j}^{\alpha \alphap}$, by construction we have $\abs{\vekrj{j}} < \abs{\vekrjp{j}}$, as illustrated in the right panel in \Figref{Fig:Scatterersjj}. Therefore, the singularity of the Green's tensor is never crossed, and we may readily expand the scalar Green's function as~\cite{MultiScat}
\begin{align} \label{Eqn:BIntgB}
\gB(\vekR') &=  \ci\kb \Dsum{\nu}{\mu} \psijbNN{j}{\nu}{\mu}(\vekrj{j}) \CC{\phib{\nu}{\mu}(\vekrjp{j})}.
\end{align}
Using this expansion to express the elements of the Green's tensor, we may write the elements $B_{j,j}^{\alpha \alphap}$ in closed form as
\begin{align} \label{Eqn:BGeneric}
\Big [B_{j,j}^{\alpha \alphap} \Big ]_{l,l'}^{m,m'} &= \ci \kb M_{l}^j \IB{j}{\lp} \Nj{\lp}{j}  /\left(\Njb{l}{j} \right) \nonumber \\
&\hspace{0.5cm} \times \bigg(\kd{\alpha}{\alphap} \kd{l}{\lp} \kd{m}{\mpr} + \frac{1}{\kb^2} \sum_{\gamma_{\alpha,\alphap}} g_{\gamma_{\alpha,\alphap}} \nonumber \\
&\hspace{0.9cm}\times \kd{l-\gamma'_{\alpha,\alphap}}{\lp} \kd{m-\gamma''_{\alpha,\alphap}}{\mpr} \bigg),
\end{align}
where $\IB{j}{\lp}$ is defined and expressed in Appendix~\ref{App:RadIntegrals}. We refer to Appendix~\ref{App:CarDerivSphWave} for details of the sum $\sum_{\gamma_{\alpha,\alphap}} g_{\gamma_{\alpha,\alphap}}$.

\subsection{Field Outside Scatterers}\label{Sec:FieldOutside}
Having determined the electric field inside the scattering objects, the Lippmann-Schwinger equation is explicit for the field at positions outside the scatterers for which we find
\begin{subequations} \label{Eqn:FieldOutsideGen}
\begin{align}
\vekE^{\alpha}(\vekr) &= \vekEB^{\alpha}(\vekr) + k_0^2 \sum_{j = 1}^N \deps_{j} \sum_{\alphap} H_{j}^{\alpha \alphap}(\vekr), \label{Eqn:FieldOutside} \\
H_{j}^{\alpha \alphap}(\vekr) &\equiv \int_{V_{j}} \GB^{\alpha \alphap}(\vekr,\vekrjp{j}) \vekE^{\alphap}(\vekrjp{j}) \ddd \vekrjp{j}. \label{Eqn:HIntegrals}
\end{align}
\end{subequations}
The integrals $H_{j}^{\alpha \alphap}(\vekr)$ can be evaluated analytically following a procedure similar to the evaluation of $B_{j,j}^{\alpha \alphap}$, as discussed in Section~\ref{Sec:SelfTerms}. 

\subsection{Background Field} \label{Sec:BackgroundField}
To solve \Eqref{Eqn:LSVecEqn} for $\vekn{a}$ we need the expansion coefficients of the background field, $\vekn{a}_{\mathrm{B}}$, cf. \Eqref{Eqn:FieldBExpan}. In the following sections, we list these coefficients for different types of excitations.
\subsubsection{Plane Wave}
We consider an incoming plane wave of the form
\begin{align} \label{Eqn:BackgroundPW}
\vekEB(\vekr) = E_0  \exp(\ci \vekk_{\mathrm{B}} \cdot \vekr) \, \vekeb,
\end{align}
where $\vekk_{\mathrm{B}}$ and $\vekeb$ are the wave vector and the unit polarization vector, respectively, satisfying $\vekk_{\mathrm{B}} \cdot \vekeb = 0$. The expansion coefficients of this field on the background spherical wavefunctions are~\cite{MultiScat}
\begin{align}
a_{j \alpha l m}^{\mathrm{B}} = \left[ E_0 \vekeb^{\alpha} 4 \pi \ci^l \lbrace \Yj{l}{m}(\theta_k,\phi_k) \rbrace^* / \left(\Njb{l}{j} \right) \right] \exp(\ci \vekk_{\mathrm{B}} \cdot \Cen{j}),
\end{align}
where $\theta_k$ and $\phi_k$ are the polar and azimuthal angles of $\vekk_{\mathrm{B}}$, respectively.\\

\subsubsection{Dipole Emitter -- Background Green's Tensor} \label{Sec:GBExpanCoefs}
The Green's tensor $\vekn{G}(\vekr,\vekr')$ is proportional to the electric field at the point $\vekr$ produced by three dipoles, with dipole moments along $\alpha \in \lbrace x,y,z \rbrace$, positioned at $\vekr'$. Consequently, we determine $\vekn{G}(\vekr,\vekr')$ by using the homogeneous background medium Green's tensor $\GB(\vekr,\vekr')$ as the background field in the Lippmann-Schwinger equation~\cite{Martin1}. To that end, we need an expansion of the background Green's tensor on the background basis functions. As shown in Appendix~\ref{App:DipoleExpanCoefs}, the expansion coefficients are
\begin{align} \label{Eqn:GBExpanCoefs}
a_{j \alpha l m}^{\mathrm{B}} &= \ci \kb (-1)^{m} / \left(\Njb{l}{j} \right) \bigg(\kd{\alpha}{\alphap} \phib{l}{-m}(\vekrjp{j}) \nonumber \\
&\hspace{0.5cm} + \frac{1}{\kb^2} \sum_{\gamma_{\alpha,\alphap}} g_{\gamma_{\alpha,\alphap}} \phib{l(\gamma_{\alpha,\alphap})}{m^*(\gamma_{\alpha,\alphap})}(\vekrjp{j}) \bigg).
\end{align}
From $\vekn{G}(\vekr,\vekr')$, the projected local density of states (LDOS) of the dipole emitter may readily be obtained as~\cite{NovotnyFull}
\begin{align} \label{Eqn:LDOS}
\rho^{\alpha}(\vekr;\lambda_0) = \frac{4}{c\lambda_0} \mathrm{Im}\left(\vekn{G}^{\alpha \alpha}(\vekr,\vekr) \right).
\end{align}
Since, in general, $\rho^{\alpha}(\vekr;\lambda_0)$ is different for different orientations of the emitter, $\alpha$, \Eqref{Eqn:LDOS} is in fact a projected LDOS. Nevertheless in the remainder of the article we shall refer to \Eqref{Eqn:LDOS} simply as the LDOS. The LDOS gives the number of modes per unit volume and frequency that the emitter can decay spontaneously into, and an important figure of merit is therefore the relative enhancement of the LDOS, $\rho^{\alpha}(\vekr;\lambda_0) / \rho_{\mathrm{B}}^{\alpha}(\lambda_0)$, due to the presence of the scattering objects. The relative LDOS is known also as the Purcell factor $F_{\mathrm{P}}$~\cite{Purcell}. As an example, we compute this quantity for a plasmonic dimer in Section~\ref{Sec:LDOSCalc}. Assuming that emissions from the individual dipoles are independent, we may include several dipole emitters at distinct positions by summing over the expansion coefficients in \Eqref{Eqn:GBExpanCoefs}, with different $\vekrjp{j}$ for each dipole emitter. An example of such a calculation, with a single spherical scattering object, is presented in~\cite{Pustovit2010}. In this case, however, the resulting field does not represent the Green's tensor.

\subsubsection{No Background Field - Quasinormal Modes} \label{Sec:QuasinormalModes}
The formalism of this work treats open systems where no boundaries enclose the structures. The modes of such open systems are inherently leaky and appear as solutions of non-Hermitian eigenvalue problems that give rise to complex eigenfrequencies $\omegat_i$. These modes are referred to as quasinormal modes~\cite{QNMMath}. The quasinormal modes $\quasi_i(\vekr;\omegat_i)$ may be determined as self-consistent solutions of an "excitation-free" Lippmann-Schwinger equation~\cite{ModeVolume}
\begin{align} \label{Eqn:QuasiLS}
\quasi_i(\vekr;\omegat_i) = k_0^2(\omegat_i)\int_{V} \GB(\vekr,\vekr';\omegat_i) \deps(\vekr';\omegat_i)  \quasi_i(\vekr';\omegat_i) \ddd \vekr'.
\end{align}
In the above, we have absorbed the technical detail of the source dyadic in the integral for brevity. Likewise, we have written the frequency dependence explicitly, and using the expansion technique developed in the previous sections and an iteration of the eigenfrequencies, we determine the quasinormal modes by solving \Eqref{Eqn:QuasiLS} self-consistently. Due to their leaky nature, the quasinormal modes give rise to finite $Q$-factors~\cite{NovotnyFull} that we can calculate as $Q_i = -\mathrm{Re}(\omegat_i) /( 2\mathrm{Im}(\omegat_i))$.

\subsection{Summary of Formalism}\label{Sec:SummaryFormalism}
We summarize the steps in the procedure for solving \Eqref{Eqn:LSEqnPrin}:
\begin{enumerate}
\item We expand each component of the total electric field $\vekE(\vekr)$ and the incoming field $\vekEB(\vekr)$ inside each of the $N$ scatterers on orthonormal sets of basis functions in \Eqsref{Eqn:FieldExpansions}.
\item The expansions are inserted into \Eqref{Eqn:LSEqnPrin}, projected onto an arbitrary basis function and summed over all free indices to yield the matrix equation for the expansion coefficients of the total electric field $\vekn{a}$ in \Eqref{Eqn:LSVecEqn}.
\item The matrices in \Eqref{Eqn:LSVecEqn} are calculated: The diagonal matrices $\MB$ and $\depsm$ are constructed directly by means of the orthonormality of the basis functions in \Eqsref{Eqn:IP}. The non-diagonal matrix $\Gm$ is evaluated using a number of expansions of the Green's tensor as discussed in Section~\ref{Sec:GreenMatElem}.
\item The incoming field is expanded. Examples for several types of excitation fields are given in Section~\ref{Sec:BackgroundField}.
\item We finally solve \Eqref{Eqn:LSVecEqn} for the expansion coefficients $\vekn{a}$.
\end{enumerate}
Subsequently, depending on the choice of incoming field, derived quantities at positions outside the scatterers may be calculated straightforwardly using \Eqsref{Eqn:FieldOutsideGen}.

\subsection{Far-Field Radiation Pattern and Extinction Cross Section}
The scattered field, i.e. the second term on the right hand side of \Eqref{Eqn:LSEqn}, must satisfy the radiation BC in the far-field~\cite{Jackson},
\begin{align} \label{Eqn:RadCond}
\vekE_{\mathrm{scat}}(\vekr) \sim \vekf(\theta,\phi) E_0 \frac{\exp(\ci \kb r)}{r}, \hspace{0.5cm} 1 \ll \kb r,
\end{align}
where $\vekf(\theta,\phi)$ is the far-field radiation pattern at the polar and azimuthal angles, $\theta$ and $\phi$, of $\vekr$. The scattered field outside the scattering objects is represented by the second term on the right hand side of \Eqref{Eqn:FieldOutside}, and these terms are all proportional to a spherical Hankel function of the first kind $\hj{l}(\kb r)$. Expanding these functions asymptotically as~\cite{Schaums}
\begin{align} \label{Eqn:SphHankelAsymp}
\hj{l}(\kb r) \sim \frac{(-\ci)^{l+1}}{\kb} \frac{\exp(\ci \kb r)}{r}, \hspace{0.5cm} 1 \ll \kb r,
\end{align}
and using the definition of the far-field radiation pattern in \Eqref{Eqn:RadCond}, we may write $\vekf(\theta,\phi)$ analytically. 

The extinction cross section is defined as the power removed from an incoming plane wave, $P$, relative to the magnitude of the incoming Poynting vector, $\Cext \equiv P / \abs{\vekSB}$. $P$ can be computed by brute force evaluation of the energy flux through a sphere enclosing the scattering objects, but may be more elegantly expressed using the Optical Theorem~\cite{Jackson,BornWolf} as
\begin{align} \label{Eqn:CextGen}
\Cext = \frac{4 \pi}{\kb} \mathrm{Im} \left(\vekf(\theta_k,\phi_k) \cdot \vekeb^* \right).
\end{align}
The extinction cross section $\Cext$ has the dimension of area and can be interpreted as an equivalent area over which the incoming radiation interacts with the scattering objects. It is therefore customary to normalize it to the geometric cross section, giving rise to the extinction efficiency $\Qext \equiv \Cext / (N\pi R^2)$, where $R$ is the radius of each of the $N$ spheres.

\subsection{Error Estimate} \label{Sec:ErrorEstimate}
As discussed in~\cite{Philip2}, the Lippmann-Schwinger equation provides a direct error estimate of the computed field. Rearranging \Eqref{Eqn:LSEqnPrin}, we may define the local error at a point $\vekr$ inside a scattering object as
\begin{align}
\EpsL(\vekr) \equiv &  \bigg | \vekEB(\vekr) - \vekE(\vekr) -\Lso \frac{\deps(\vekr)}{\epsb}\vekE(\vekr)  \nonumber \\
&
+ k_0^2 \int_{V_{\mathrm{scat}}-\delta V} 
\GB(\vekr,\vekr') \deps(\vekr') \vekE(\vekr') \ddd \vekr' \bigg |.
\end{align}
In this expression, $\vekE(\vekr)$ is the approximation to the electric field, i.e. \Eqref{Eqn:FieldExpan} with a finite number of basis functions retained. We also define the global relative error
\begin{align}
\EpsG \equiv \frac{\int_{V_{\mathrm{scat}}} \EpsL(\vekr) \ddd \vekr}{\int_{V_{\mathrm{scat}}} \abs{\vekE(\vekr)} \ddd \vekr},
\end{align}
which represents an explicit error estimate of the field. In practice, the set of basis functions is truncated by truncating $l$ at $\lmax$, and a specific calculation of $\EpsG$ is presented in Section~\ref{Sec:PlaneWaveScattering}. In the case of spherical scatterers, as considered in this work, all matrix elements are expressed analytically, and the truncation of the set of basis functions therefore represents the most significant approximation in the formalism.

\section{Example Calculations: Plasmonic Dimer}\label{Sec:PlasmonicDimer}
It is well-known that metallic nanoparticles may sustain collective oscillations of free charges known as plasmons. A special type of plasmons are localized surface plasmons (LSPs), bound to the interface between a metal and a dielectric, that give rise to strongly enhanced near-fields~\cite{Maier}. The plasmonic dimer that we examine in the following sections may support LSPs, and it has been demonstrated that the resonance wavelengths of these depend sensitively on parameters such as the distance between the particles and their sizes~\cite{Zhang2008}. The LSP occurring at the largest excitation wavelength is known as the dipole LSP~\cite{Myroshnychenko2008,Hao2004}, while resonances at shorter wavelengths are higher-order LSPs. It has been shown that the dipole resonance redshifts or blueshifts for decreasing distance between the metal nanoparticles for parallel or perpendicular polarization (with respect to the dimer axis) of the incoming field, respectively~\cite{3DPlasmonChain}. For parallel polarization, it has been suggested that the relative shift of the resonance wavelength for the dimer depends exponentially on the gap distance between the particles. This analysis was based on a qualitative description for the dimer as an electric dipole~\cite{PlasmonRuler}. For gap sizes below the radius of the particles, however, the strong coupling of the near-fields between the particles makes this simple description invalid~\cite{PlasmonChains}, and modeling including higher-order wavefunctions, as in the present approach, is needed to correctly analyze the closely spaced nanoparticles. 

\begin{figure}[htb!]
\centering
\begin{tikzpicture}[line cap=round,line join=round,>=triangle 45,x=1.0cm,y=1.0cm,line width=0.3mm]

\shade[ball color=blue!10!white,opacity=0.20] (0,0) circle (1cm);
\shade[ball color=blue!10!white,opacity=0.20] (3,0) circle (1cm);

\newcommand\dx{-5}
\newcommand\dy{-1}
\draw [->] (2.5+\dx,1+\dy)  -- (3.5+\dx,1+\dy) node(xline)[right]{$y$};
\draw [->] (2.5+\dx,1+\dy)  -- (2.5+\dx,2+\dy) node(yline)[right]{$z$}; 
\draw [->] (2.5+\dx,1+\dy)  -- (1.9+\dx,0.4+\dy) node(xline)[below]{$x$}; 

\fill [color=black] (0,0) circle (2.5pt);
\fill [color=black] (3,0) circle (2.5pt);

\draw (-1,0) arc (180:360:1cm and 0.5cm);
\draw[dashed] (-1,0) arc (180:0:1cm and 0.5cm);
\draw[dashed] (0,1) arc (90:-90:0.5cm and 1cm);
\draw (0,1) arc (90:270:0.5cm and 1cm);
\draw (0,0) circle (1cm);

\draw (2,0) arc (180:360:1cm and 0.5cm);
\draw[dashed] (2,0) arc (180:0:1cm and 0.5cm);
\draw[dashed] (3,1) arc (90:-90:0.5cm and 1cm);
\draw (3,1) arc (90:270:0.5cm and 1cm);
\draw (3,0) circle (1cm);

\draw[<->,color=blue,line width=1pt] (1,0) -- (2,0);
\draw[<->,color=blue,line width=1pt] (0,0) -- (0,1);
\draw[<->,color=blue,line width=1pt] (3,0) -- (3,1);

\draw (1.5,0.3) node[circle]{$d$};
\node[anchor=west] at (0,0.5) (text) {};
\node[anchor=south] at (-2,1.3) (description) {$R=25\nm$};
\draw[->] (description) to [out = 355, in = 180] (text);
\draw (0,-0.75) node[circle]{$\epsind{Ag}$};
\draw (3,-0.75) node[circle]{$\epsind{Ag}$};

\end{tikzpicture}
\caption{(Color online) Schematic of plasmonic dimer consisting of two Ag spheres aligned along the $y$-axis. Each particle has radius $R=25\nm$, and the surface-to-surface distance between the spheres is $d$.}
\label{Fig:DimerSchematic}
\end{figure}
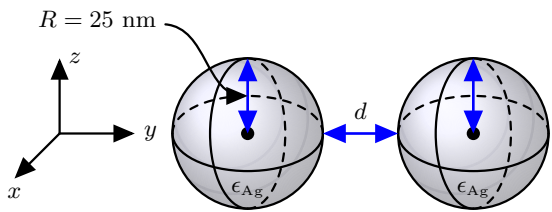

The dimer consists of two Ag particles aligned along the $y$-axis, each of radius $R = 25\nm$ and spaced a distance $d$ apart, as shown in \Figref{Fig:DimerSchematic}. In all calculations, $R=25\nm$ is fixed while $d$ is a parameter. The particles are embedded in SiO$_2$ ($\epsb = 2.25$), and the permittivity of the Ag spheres is given by the Drude model $\epsilon(\omega) = 1-\omega_{\mathrm{p}}^2 / (\omega^2 + \ci \gamma \omega)$ with $\hbar \omega_{\mathrm{p}} = 7.9$ eV and $\hbar \gamma = 0.06$ eV~\cite{Koenderink}.

\subsection{Plane Wave Scattering} \label{Sec:PlaneWaveScattering}
We illuminate the dimer by plane waves polarized parallel ($\vekeb = \hat{y}$) or perpendicularly ($\vekeb = \hat{x}$) to the dimer axis. In both cases, the incoming field propagates perpendicularly to the dimer axis ($\vekkb = \kb \hat{z}$).

\begin{figure}[htb!]
\centering
\begin{tikzpicture} [line cap=round,line join=round,>=triangle 45,x=1.0cm,y=1.0cm, scale = 0.47]
\node (img) at (0,0) {\includegraphics[scale=0.3]{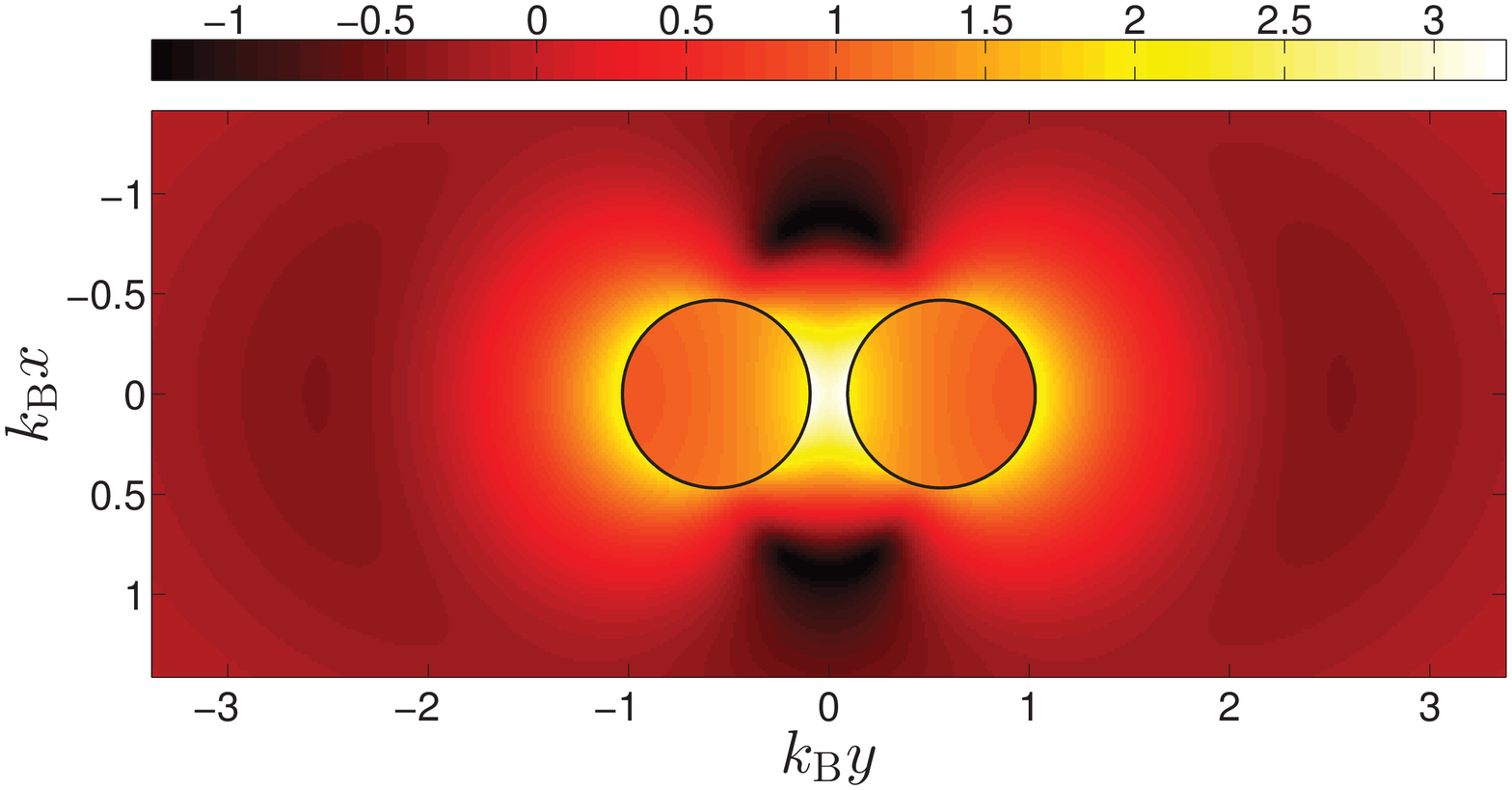}};
\node (img) at (2.5,0) {
\begin{tikzpicture}[line cap=round,line join=round,>=triangle 45,x=1.0cm,y=1.0cm,transform canvas={scale=0.47}]

\draw [<->,line width=2pt,color=black] (2,0)  -- (4.5,0); 
\draw (3.25,-0.3) node[circle]{\Large{$\vekn{e}_{\mathrm{B}}$}}; 

\end{tikzpicture}
};
\node (img) at (0,-9) {\includegraphics[scale=0.3]{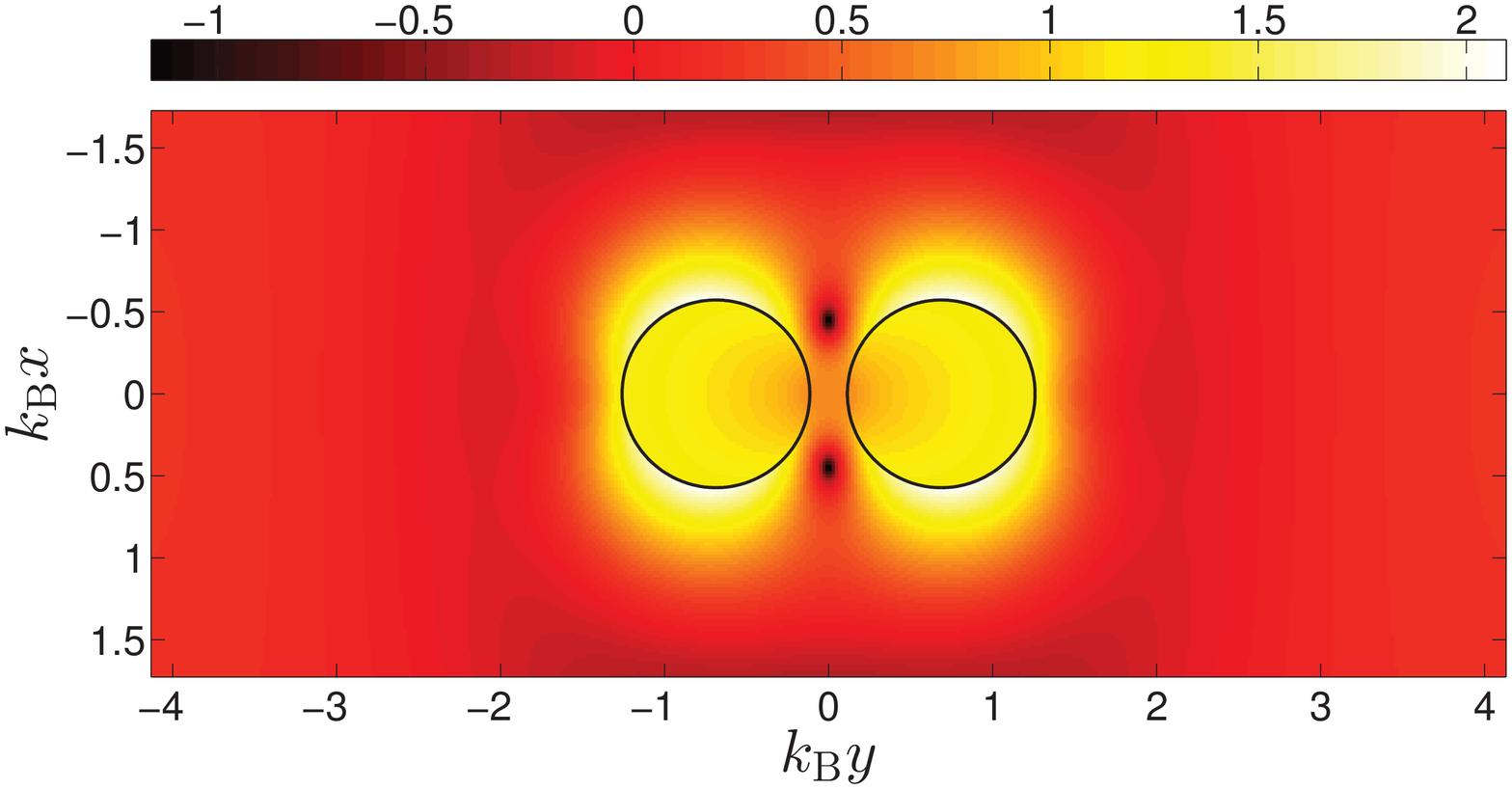}};
\node (img) at (2.5,-9) {
\begin{tikzpicture}[line cap=round,line join=round,>=triangle 45,x=1.0cm,y=1.0cm,transform canvas={scale=0.47}]

\draw [<->,line width=2pt,color=black] (3.25,-1.25)  -- (3.25,1.25); 
\draw (2.75,0) node[circle]{\Large{$\vekn{e}_{\mathrm{B}}$}}; 

\end{tikzpicture}
};
\end{tikzpicture}
\caption{(Color online) Relative enhancement of the electric field intensity, $\log_{10}\left( \abs{\vekE}^2 / \abs{\vekEB}^2\right)$, for scattering of a plane wave on a dimer, shown in the $xy$-plane at $z=0$. The spacing between the particles is $d=10\nm$. The incoming plane wave propagates along the $z$-direction and is polarized parallel (top panel, $\lambda_0 = 505 \nm$) or perpendicular (bottom panel, $\lambda_0 = 412 \nm$) to the dimer axis. Note that different color scales are used in the two plots.}
\label{Fig:FieldScatReson}
\end{figure}
Two LSPs excited by the plane waves are displayed in \Figref{Fig:FieldScatReson}, showing the relative enhancement of the electric field intensity in a log-scale, in the $xy$-plane at $z=0$. The distance between the spheres is $d=10\nm$. The maximum relative intensity enhancement is roughly an order of magnitude larger for parallel than for perpendicular polarization. For parallel polarization the enhancement occurs in the gap, while for perpendicular maximum enhancement occurs at the surfaces of the particles. This is caused by the charge oscillations in the metal particles that move in phase with the exciting field~\cite{PlasmonRuler}; In the parallel case, this produces a charge difference across the gap which gives an enhanced field between the particles, while in the perpendicular case the charge distributions give strongly enhanced fields on the individual particles.
\begin{figure}[htb!]
\centering
\includegraphics[scale=0.35]{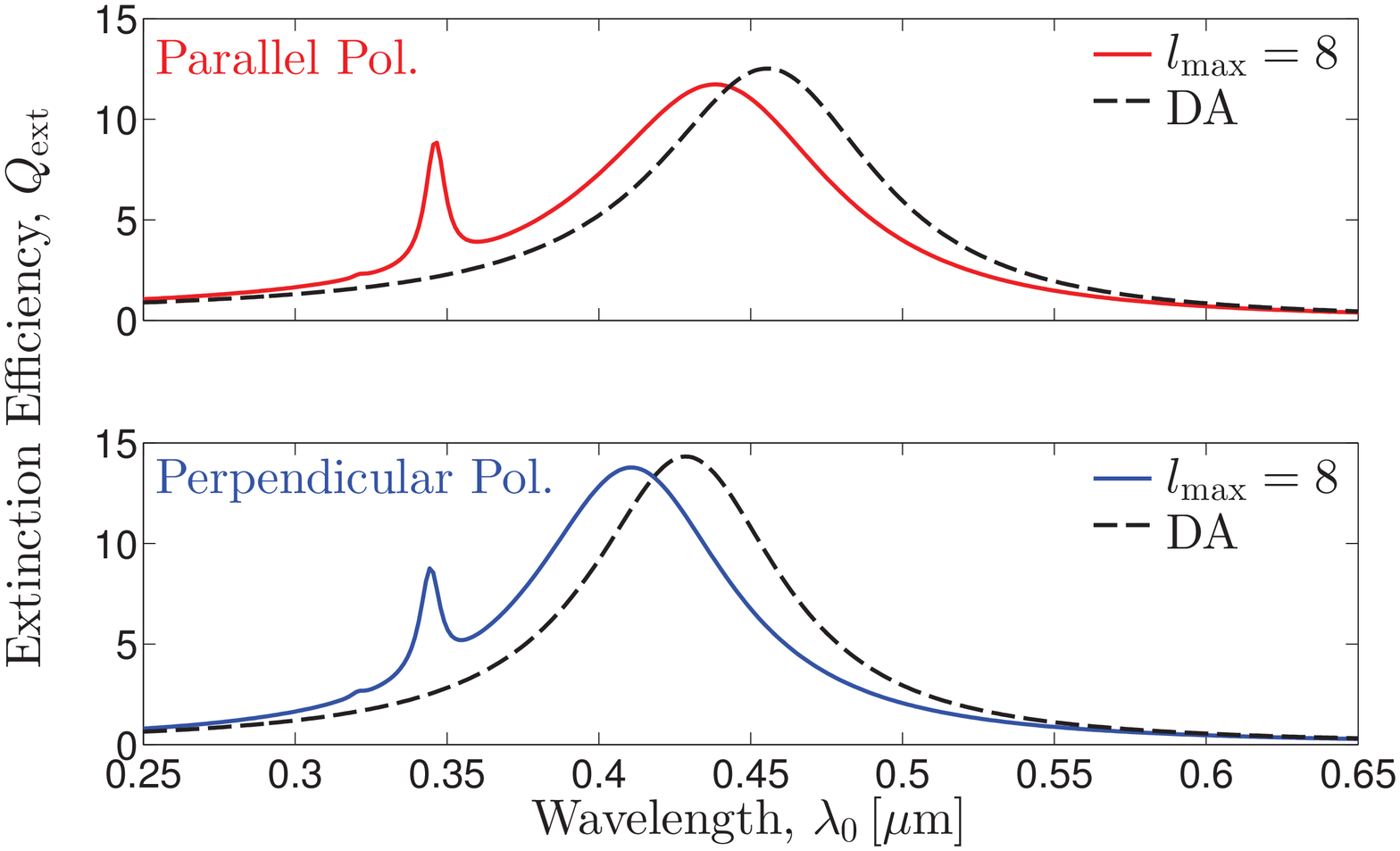}
\caption{(Color online) Extinction efficiency $\Qext$ versus excitation wavelength for scattering of a plane wave on two Ag spheres spaced by $d = 50\nm$. The incoming field is polarized parallel (top panel) or perpendicularly (bottom panel) to the dimer axis. The spectra have been obtained using the full formalism with $\lmax = 8$ and the dipole approximation (DA).}
\label{Fig:QextDimerDA25nm}
\end{figure}

\Figref{Fig:QextDimerDA25nm} shows the extinction efficiency spectrum for a dimer with spacing $d = 50\nm$ which is excited by parallel (top panel) and perpendicular (bottom panel) plane waves. The spectra are computed using the full formalism developed in Section~\ref{Sec:MutiScatFormalism} ($\lmax = 8$, full curves) and using the dipole approximation (dashed). We have also computed the spectra using $\lmax = 10$, and comparing these with the $\lmax = 8$ spectra we find a vanishing relative deviation on the dipole resonance wavelength for both polarizations and a relative deviation on the resonance values of $\sim 10^{-13}$.
The spectra obtained using the full formalism and using the DA look qualitatively the same: They predict the dipole resonances at $\lambda_0 \sim 450\nm$ and at $\lambda_0 \sim 425\nm$ in the parallel and perpendicular cases, respectively. The relative deviations of the DA-calculations for the dipole resonance wavelengths are in both cases 4\%, while the relative deviation, on resonance, of the extinction efficiency is 7\% and 4\%, respectively. These relative errors increase for decreasing spacing $d$ (not shown), illustrating the limitation of the DA for closely spaced nanoparticles. Another shortcoming of the DA is that it is inherently monomode; it only predicts the dipole resonance and not the higher order mode occurring at $\lambda_0 \sim 340 \nm$.
In line with the DA, the quasistatic approximation is another popular approximate scheme for solving \Eqref{Eqn:WaveEqn}. In~\cite{Chen2010}, Chen \textit{et al.} analyzed the electrodynamic coupling between a quantum dot and a plasmonic nanowire. In particular, it was demonstrated that the characteristic size of the scatterer (plasmonic nanowire) needs to be smaller than the skin depth of the metal for the quasistatic approximation to accurately model the coupling. The results in Fig.~\ref{Fig:QextDimerDA25nm} and in~\cite{Chen2010} demonstrate the need for full vectorial solvers for modeling nanoplasmonic structures when these contain small features, e.g. small spacings as in the example of this section.
\begin{figure}[htb!]
\centering
\includegraphics[scale=0.35]{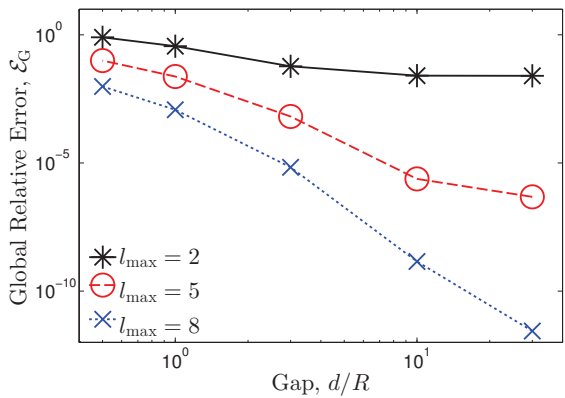}
\caption{(Color online) Global relative error of the electric field as function of distance between two Ag spheres, $d/R$. Results for three different truncations of the set of basis functions are shown. Illumination by plane waves at $\lambda_0 = 800\nm$ with oblique incidence and polarization.} \label{Fig:ErrorEstimate} 
\end{figure}

To further investigate the impact of truncating the basis set, \Figref{Fig:ErrorEstimate} shows the global relative error $\EpsG$ of the electric field for the dimer as a function of $d/R$ and for three values of $\lmax$. The system is illuminated by a plane wave as detailed in the caption of the figure. At the largest spacing, $d/R = 30$, the global relative error decreases by five to six orders of magnitude when $\lmax$ increases by three units. As $d/R$ decreases, the global relative error increases for fixed $\lmax$. This reduction in accuracy is caused by the introduction of an increasingly smaller length scale, namely the distance between the particles, and more basis functions are needed to resolve this correctly. As an example, with $\lmax = 2$ and for $d/R = 1$ and $d/R = 0.5$ the global relative errors equal 36\% and 80\%, respectively, which supports the conclusion that vectorial solvers are needed to correctly model closely spaced nanoparticles. Similar results were obtained using a different solution technique in~\cite{PlasmonChains}. We note, that in the limit $d/R \ll 1$, nonlocal effects~\cite{Raza} become important and must be included to correctly model the field. For fixed particle distance, the exponential decay of the relative error in \Figref{Fig:ErrorEstimate} enables highly accurate calculations with a modest number of basis functions, and in the following sections we use $\lmax = 8$.

\subsection{Dipole Emitter and LDOS} \label{Sec:LDOSCalc}
\begin{figure}[htb!]
\centering
\includegraphics[scale=0.3]{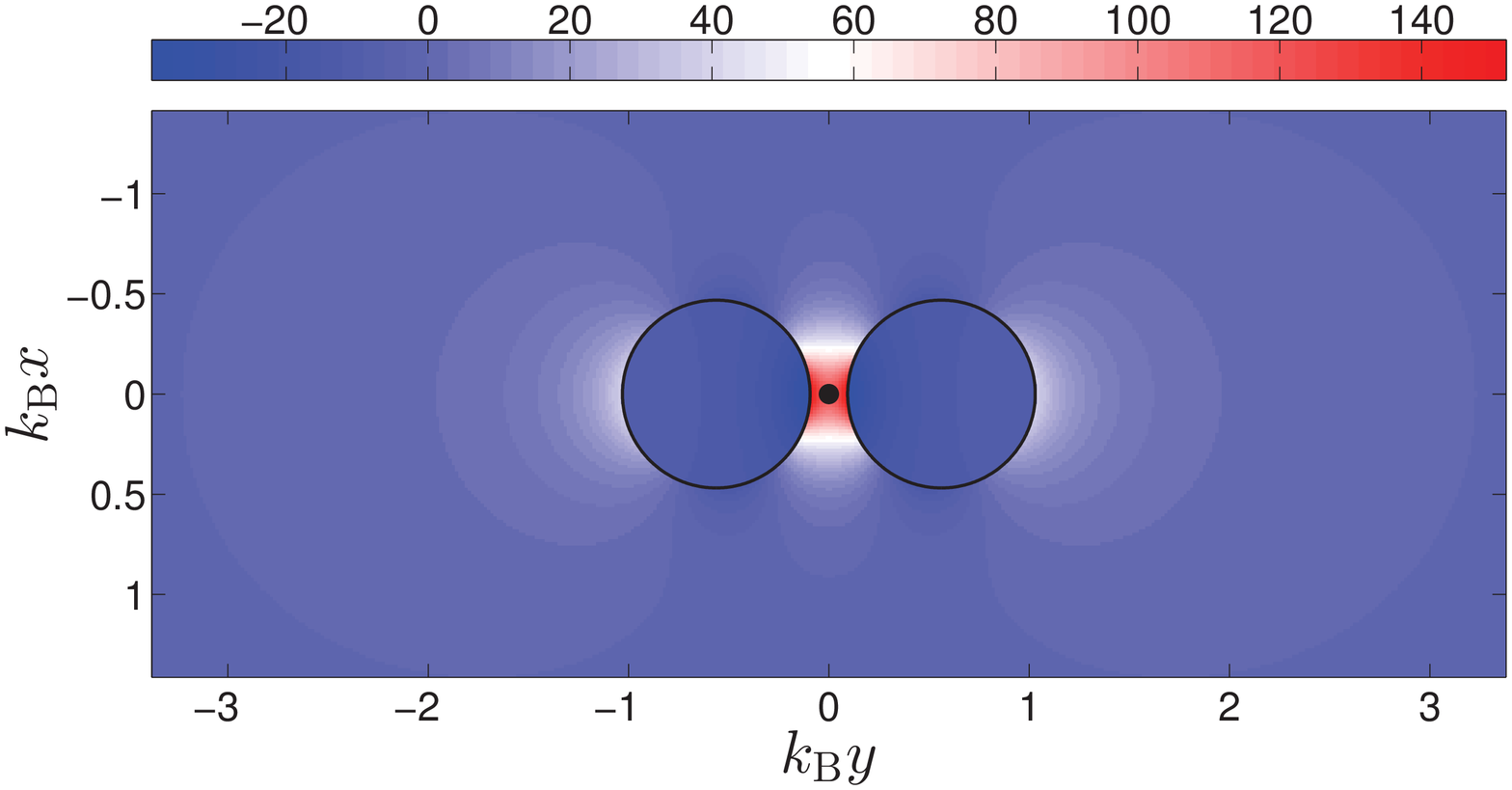}
\includegraphics[scale=0.3]{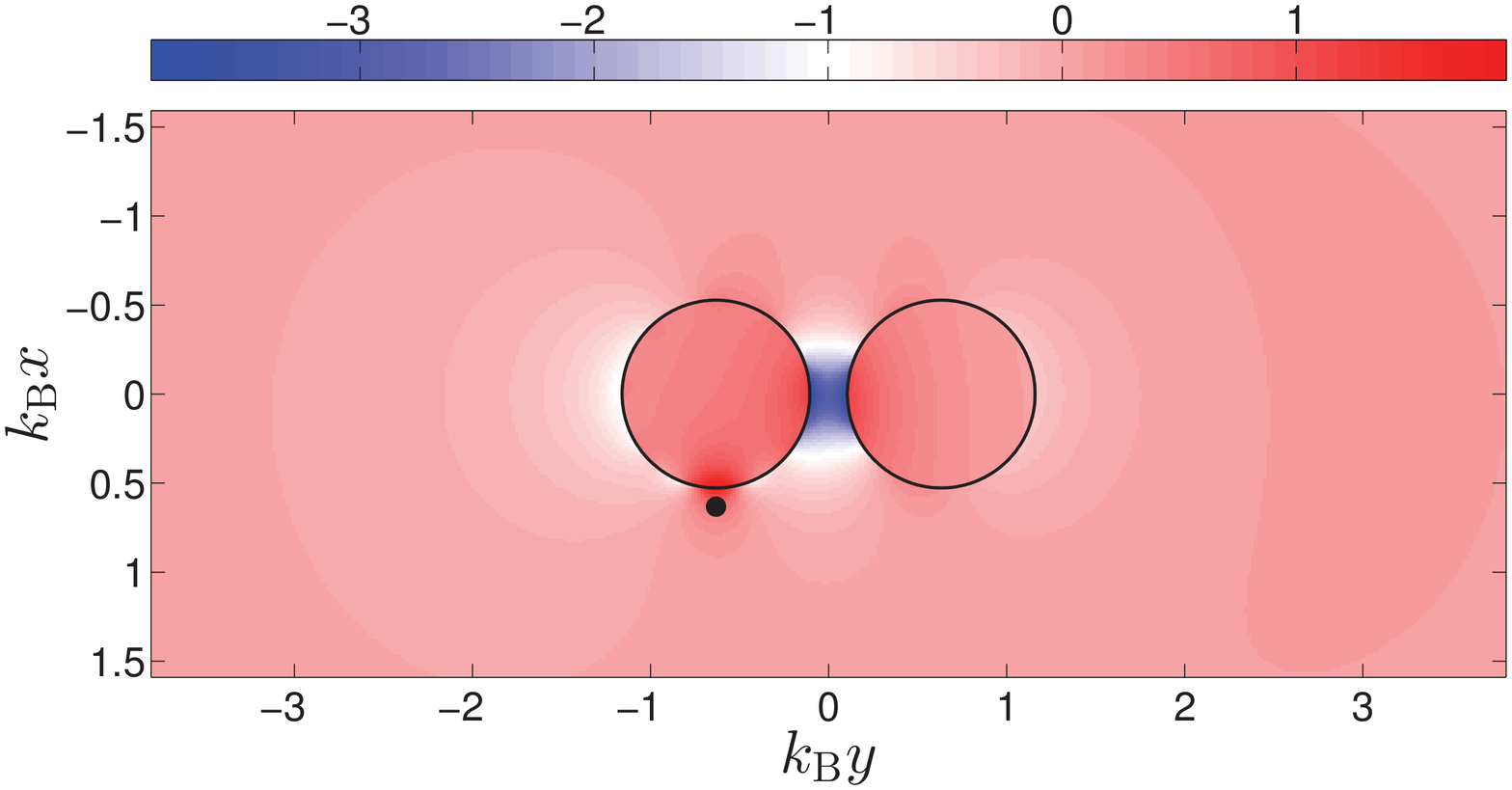}
\caption{(Color online) Imaginary part of Green's tensor $\mathrm{Im}(\vekn{G}^{yy}(\vekr,\vekr'))$ relative to $\mathrm{Im}(\GB^{yy}(\vekr',\vekr'))$ for the dimer with two Ag spheres ($d/R = 0.4$). The two panels show results with different positions of the dipole emitter, $\vekr'$, (black dots) and different wavelengths $\lambda_0 = 505\nm$ (top) and $\lambda_0 = 447\nm$ (bottom). Note that different color scales are used in the two plots.}
\label{Fig:Greens_Imagy}
\end{figure}

We embed a dipole emitter with dipole moment along the direction $\alpha$ in the vicinity of the plasmonic dimer. The field radiated by the emitter equals the $\alpha$th column of the Green's tensor $\vekn{G}(\vekr,\vekr')$ where $\vekr'$ is the dipole position. \Figref{Fig:Greens_Imagy} shows the imaginary part of the Green's tensor $\mathrm{Im}(\vekn{G}^{yy}(\vekr,\vekr'))$ relative to $\mathrm{Im}(\GB^{yy}(\vekr',\vekr'))$ for two positions of the dipole $\vekr'$ (indicated by dots in the figure) and at two different wavelengths, $\lambda_0 = 505\nm$ (top panel) and $\lambda_0 = 447\nm$ (bottom panel).
In the top panel, the imaginary part of the Green's tensor is strongly enhanced relative to the background Green's tensor at the position of the emitter. Conversely, the imaginary part of the Green's tensor is not enhanced at the emitter position in the bottom panel. From the expression for the LDOS in \Eqref{Eqn:LDOS} we know that the LDOS is proportional to the imaginary part of the Green's tensor at the emitter position. To further investigate the emission properties of the dipole emitter, we analyze the Purcell factor in the following.
\begin{figure}[htb!]
\centering
\begin{tikzpicture} [line cap=round,line join=round,>=triangle 45,x=1.0cm,y=1.0cm, scale = 0.47]
\node (img) at (0,0) {\includegraphics[scale=0.35]{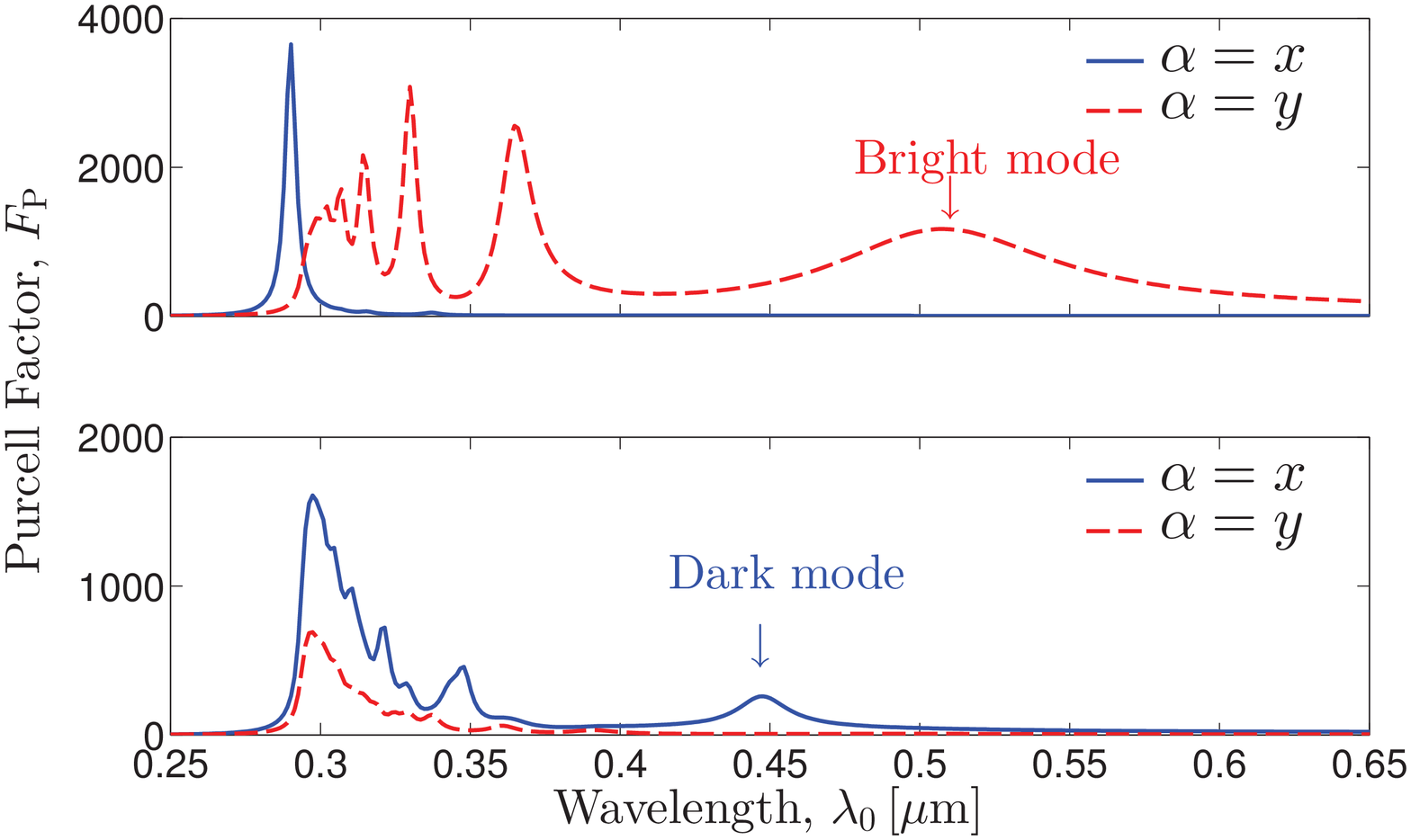}};
\node (img) at (-1.3,3.9) {
\begin{tikzpicture}[line cap=round,line join=round,>=triangle 45,x=1.0cm,y=1.0cm,transform canvas={scale=0.24},font=\Huge,line width=0.3mm]

\fill [color=black] (2-0.4,-1) circle (2.5pt); 
\draw [->,-latex,line width=1pt] (2-0.4,-1)  -- (3-0.4,-1) node(yline)[right]{$y$}; 
\draw [->,-latex,line width=1pt] (2-0.4,-1)  -- (2-0.4,-2) node(xline)[right]{$x$}; 

\draw (0,0) circle (1cm);
\shade[ball color=blue!10!white,opacity=0.20] (0,0) circle (1cm);

\draw (4,0) circle (1cm);
\shade[ball color=blue!10!white,opacity=0.20] (4,0) circle (1cm);
 
\fill [color=black] (2,0) circle (3.5pt);
\draw (2,0.5) node[circle]{$\vekr_1$};

\end{tikzpicture}
};
\node (img) at (-3.2,-1.5) {
\begin{tikzpicture}[line cap=round,line join=round,>=triangle 45,x=1.0cm,y=1.0cm,transform canvas={scale=0.24},font=\Huge,line width=0.3mm]

\fill [color=black] (2-0.4,-1) circle (2.5pt); 
\draw [->,-latex,line width=1pt] (2-0.4,-1)  -- (3-0.4,-1) node(yline)[right]{$y$}; 
\draw [->,-latex,line width=1pt] (2-0.4,-1)  -- (2-0.4,-2) node(xline)[right]{$x$}; 

\draw (0,0) circle (1cm);
\shade[ball color=blue!10!white,opacity=0.20] (0,0) circle (1cm);

\draw (4,0) circle (1cm);
\shade[ball color=blue!10!white,opacity=0.20] (4,0) circle (1cm);

\fill [color=black] (0,-1.85) circle (3.5pt);
\draw (0,-2.5) node[circle]{$\vekr_2$}; 

\end{tikzpicture}
};
\end{tikzpicture}
\caption{(Color online) Spectra of Purcell factor for dipole emitter at two positions, $\vekr_1$ (top panel) and $\vekr_2$ (bottom panel), in the $z=0$-plane in the vicinity of Ag dimer with $d/R = 0.4$. Two orientations of the dipole moment of the emitter, $\alpha \in \lbrace x,y \rbrace$, have been employed. Note that different scalings are used in the two plots.} \label{Fig:LDOSDimer} 
\end{figure}

We compute the Purcell factor for two orientations of the dipole moment, $\alpha \in \lbrace x,y \rbrace$, and at the two positions from \Figref{Fig:Greens_Imagy}, producing the spectra in \Figref{Fig:LDOSDimer}. At the symmetric position $\vekr = \vekr_1$ (top panel), the Purcell factor has a single peak at $\lambda_0 \sim 290\nm$ for an $x$-oriented dipole (full curve). In contrast, for a dipole moment along the dimer axis, $\alpha = y$ (dashed curve), the Purcell factor is larger than unity over most of the considered wavelength range, $280 \nm \lesssim \lambda_0$, and exhibits a maximum at $\lambda_0\sim 505 \nm$, resulting in the enhancement that was noted in the discussion of the top panel in \Figref{Fig:Greens_Imagy}. At the asymmetric position $\vekr = \vekr_2$ (bottom panel), the Purcell factor is generally smaller than at $\vekr = \vekr_1$. For $x$-orientation of the dipole (full curve), the Purcell factor is larger than unity in the range $290 \nm \lesssim \lambda_0 \lesssim 500\nm$, and exhibits multiple peaks in this range. For a dipole moment along the dimer axis, $\alpha = y$ (dashed curve), the Purcell factor is larger than one in a narrow bandwidth, $300 \nm \lesssim \lambda_0 \lesssim 350 \nm$, but is otherwise suppressed, in particular at $\lambda_0 \sim 447\nm$ in agreement with the discussion of the bottom panel in \Figref{Fig:Greens_Imagy}.
In conclusion, the results in \Figref{Fig:LDOSDimer} show a rich display of peaks, each of which we can associate with a quasinormal mode of the dimer. In the following section, we focus on two of these as noted in \Figref{Fig:LDOSDimer}.

\subsection{Bright and Dark Modes}
The dimer introduced in \Figref{Fig:LDOSDimer} was analyzed by Koenderink in~\cite{Koenderink}, using the method of~\cite{GarciadeAbajo1999}. In particular, Koenderink found a mode of that structure with a quality factor of $Q = 5.7$ and at $\lambda_0 \sim 506\nm$. However, no rigorous definition of the mode was given. The plasmonic dimer supports both bright and dark quasinormal modes; the former can be excited both from the far-field (plane waves) and the near-field (dipole emitter), while the latter can only be excited from the near-field~\cite{DarkBrightModes}. Each of the peaks in the Purcell factor spectra correspond to the existence of a quasinormal mode. For the dimer of \Figref{Fig:LDOSDimer}, we find two of the lowest order modes at $\mathrm{Re}(\lambda_0^{\mathrm{bright}}) = 505\nm$ and $\mathrm{Re}(\lambda_0^{\mathrm{dark}}) = 447\nm$, as indicated in the figure, with low quality factors of $Q^{\mathrm{bright}} = 5.7$ and $Q^{\mathrm{dark}} = 22.1$, respectively. The bright mode is the mode that Koenderink found, demonstrating a quantitative agreement between our method and that of~\cite{GarciadeAbajo1999}.
\begin{figure}[htb!]
\centering
\begin{tikzpicture} [line cap=round,line join=round,>=triangle 45,x=1.0cm,y=1.0cm, scale = 0.47]
\node (img) at (4,0) {\includegraphics[scale=0.3]{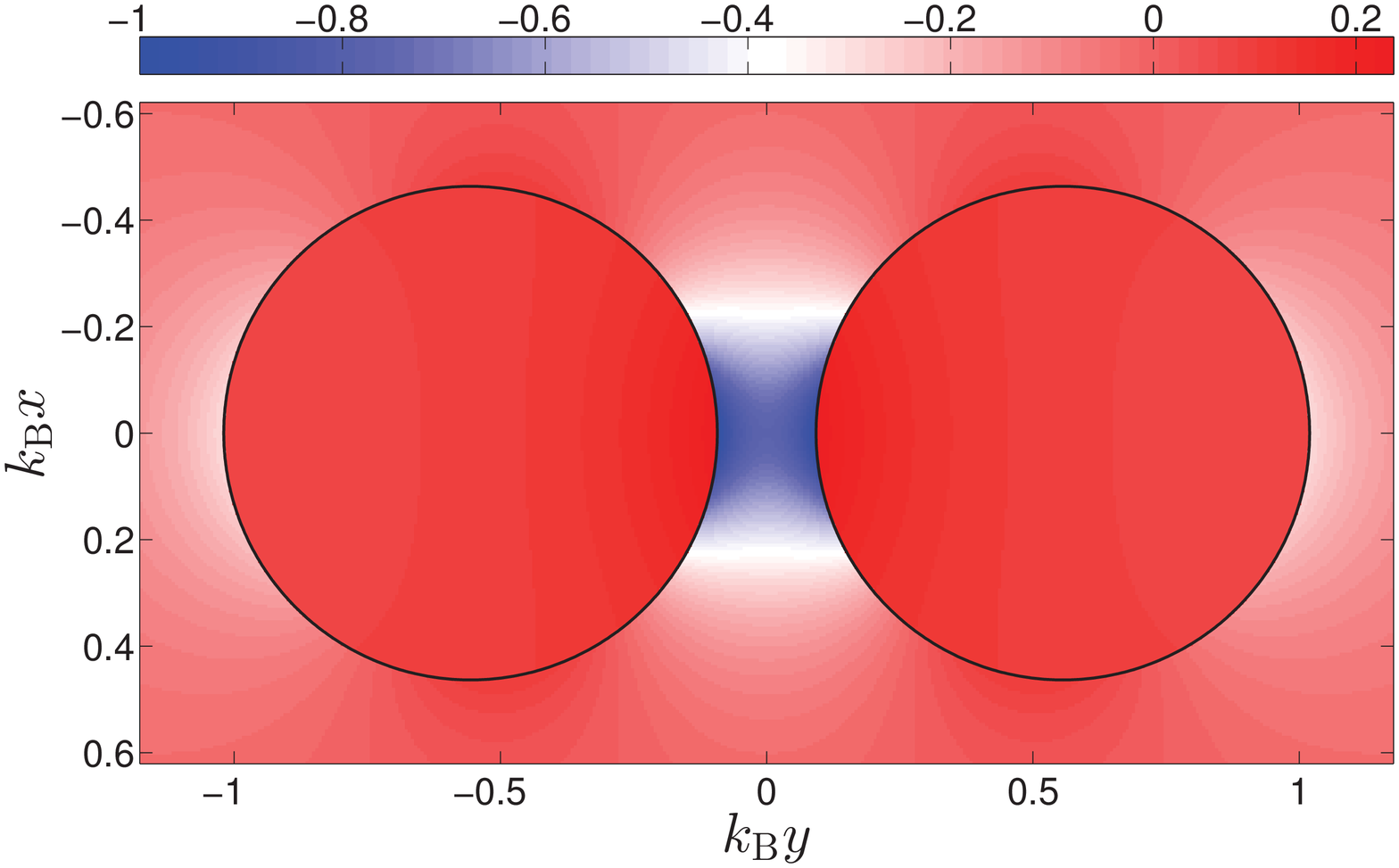}};
\node (img) at (4.7,0) {
\begin{tikzpicture}[line cap=round,line join=round,>=triangle 45,x=1.0cm,y=1.0cm]
\newcommand{\dxL}{0.75}
\newcommand{\dxR}{0.6}

\draw (0.8+\dxL,0) node[circle]{$-$}; 
\draw (0.75+\dxL,0.35) node[circle]{$-$}; 
\draw (0.75+\dxL,-0.35) node[circle]{$-$}; 
\draw (0.6+\dxL,-0.6) node[circle]{$-$}; 
\draw (0.6+\dxL,0.6) node[circle]{$-$}; 

\draw (3.2-\dxR,0) node[circle]{$+$}; 
\draw (3.25-\dxR,0.35) node[circle]{$+$}; 
\draw (3.25-\dxR,-0.35) node[circle]{$+$}; 
\draw (3.4-\dxR,-0.6) node[circle]{$+$}; 
\draw (3.4-\dxR,0.6) node[circle]{$+$}; 
\end{tikzpicture}
};
\node (img) at (4,-12) {\includegraphics[scale=0.3]{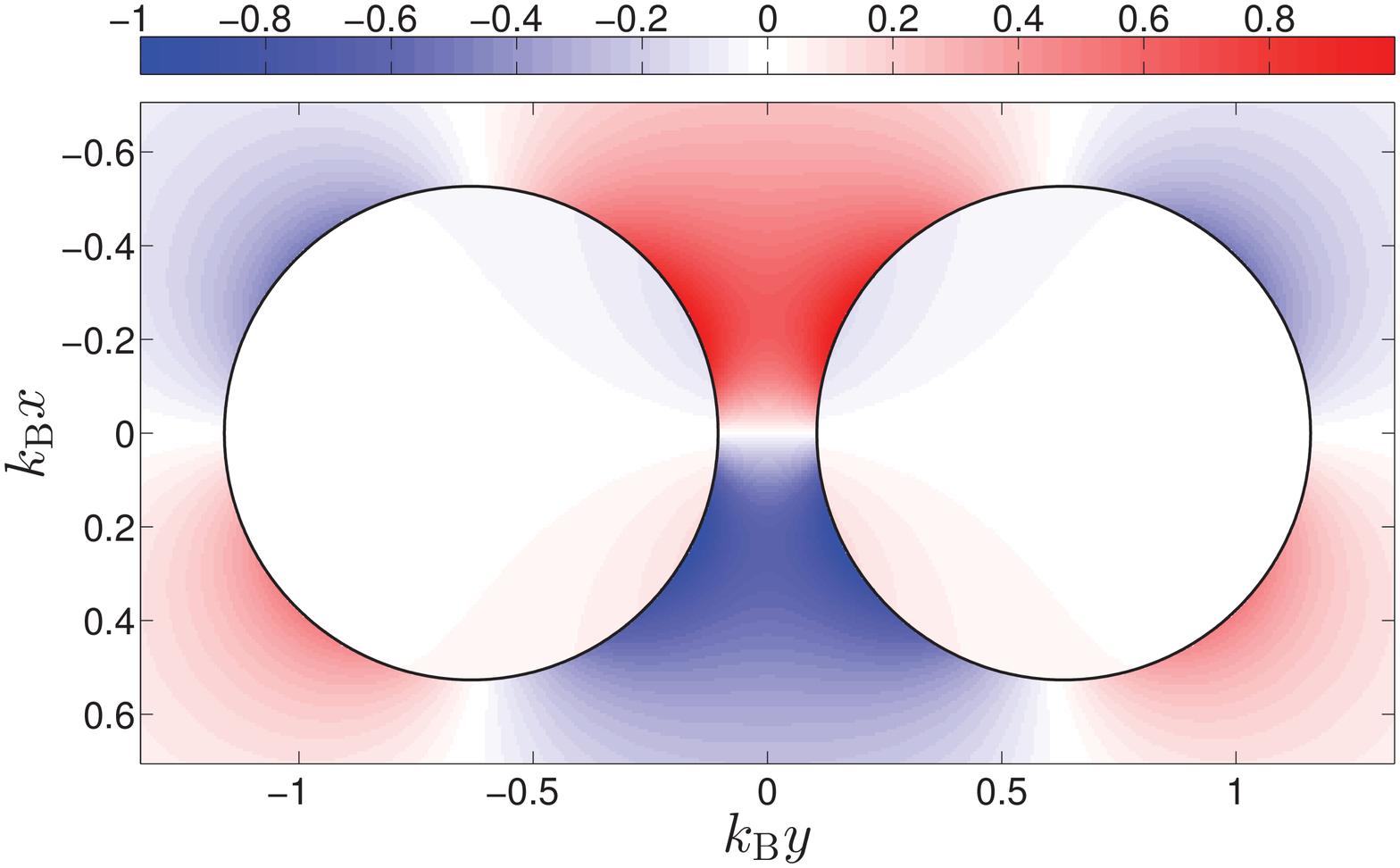}};
\node (img) at (4.9,-12.1) {
\begin{tikzpicture}[line cap=round,line join=round,>=triangle 45,x=1.0cm,y=1.0cm]
\newcommand{\dxL}{1.3}
\newcommand{\dxR}{0}
\newcommand{\dy}{0}

\draw (0.85+\dxL,0-0.4+\dy) node[circle]{$+$};
\draw (0.75+\dxL,0.35-0.4+\dy) node[circle]{$+$}; 
\draw (0.6+\dxL,0.6-0.4+\dy) node[circle]{$+$};
\draw (0.40+\dxL,0.75-0.4+\dy) node[circle]{$+$};
\draw (0.85+\dxL,-1-0.2+\dy) node[circle]{$-$};
\draw (0.75+\dxL,-1.35-0.2+\dy) node[circle]{$-$}; 
\draw (0.6+\dxL,-1.6-0.2+\dy) node[circle]{$-$}; 
\draw (0.40+\dxL,-1.75-0.2+\dy) node[circle]{$-$};

\draw (3.4+\dxR,0-0.4+\dy) node[circle]{$-$};
\draw (3.6+\dxR,0.35-0.4+\dy) node[circle]{$-$}; 
\draw (3.75+\dxR,0.6-0.4+\dy) node[circle]{$-$};
\draw (4+\dxR,0.75-0.4+\dy) node[circle]{$-$};
\draw (3.4+\dxR,-1-0.2+\dy) node[circle]{$+$};
\draw (3.6+\dxR,-1.35-0.2+\dy) node[circle]{$+$}; 
\draw (3.75+\dxR,-1.6-0.2+\dy) node[circle]{$+$}; 
\draw (4.1+\dxR,-1.75-0.2+\dy) node[circle]{$+$};

\end{tikzpicture}
};
\end{tikzpicture}
\caption{(Color online) Real part of $y$-components of quasinormal modes of plasmonic dimer, as specified in \Figref{Fig:LDOSDimer}, in the $xy$-plane. The modes, found at $\mathrm{Re}(\lambda_0^{\mathrm{bright}}) = 505\nm$ (top panel) and $\mathrm{Re}(\lambda_0^{\mathrm{dark}}) = 447\nm$ (bottom panel), respectively, are bright and dark modes, respectively. Note that different scalings are used in the two plots.} \label{Fig:QuasinormalModes} 
\end{figure}

\Figref{Fig:QuasinormalModes} shows the real parts of the $y$-components of the two modes in the $xy$-plane; the top and the bottom panels show the bright and the dark mode, respectively. In the two cases, the surface charge distributions that act to sustain the modes are indicated schematically. The bright mode is uniform in the gap between the particles due to the charge difference across the gap. This charge distribution gives rise to a finite dipole moment of the dimer which is excitable by incoming plane waves. In contrast, due to the asymmetry in the charge distribution the dark mode changes sign inside the gap, as shown in the figure. This charge distribution implies a zero net dipole moment of the mode, and consequently it is \textit{not} excitable by plane waves.

\section{Conclusion} \label{Sec:Conclusion}
We have developed a formalism for solving the Lippmann-Schwinger equation for the electric field in a 3D background medium with inhomogeneities. We have in detail explained the solution procedure for the specific example of spherical scattering objects, and based on expansions in spherical wavefunctions and addition and translation theorems we have expressed all parts of the formalism analytically. We stress that other shapes of the scatterers as well as inhomogeneous background media can be treated within the presented framework, although it may not be possible to express the matrix elements analytically. We have presented an explicit error estimate for the calculated electric field, and illustrated how this is an important tool for assessing the accuracy of the calculations. Using the formalism, we have shown how to calculate a number of physically important quantities, including the far-field radiation pattern, the extinction cross section, the total Green's tensor, the projected local density of states, the Purcell factor and quasinormal modes with their associated $Q$-factors. In particular, the analysis of LSPs as quasinormal modes provides additional physical insights, and we believe that more complex structures can benefit from this modal approach. Similar analyses of more complicated systems such as larger arrays of plasmonic nanoparticles or photonic crystals can readily be carried out; see~\cite{deLasson2012} for an example.

We believe that the versatility of the formalism will be useful for a variety of computational problems within nanophotonics. The formulation may benefit from known expressions for the background Green's tensor for different background media, e.g. for modeling of layered 3D structures~\cite{LayeredGreens}. As another perspective, we mention that the present formalism may be used for simulating electron energy loss spectroscopy (EELS) on plasmonic nanospheres. The incoming electron beam in EELS gives rise to a background electric field, and expanding this field on the background basis functions~\cite{GarciadeAbajo1999}, the EELS-field may be straightforwardly determined with the formalism developed in this paper.

\appendix
\section{Spherical Wavefunctions}\label{App:SphWaveFunctions}
The basis functions used in the expansions in \Eqsref{Eqn:FieldExpansions} are defined as follows
\begin{subequations}
\begin{align}
\psij{j}{l}{m}(\vekrj{j}) &\equiv S_j(\vekr) \Nj{l}{j} \jj{l}(\kj{j} \rj{j}) \Yj{l}{m}(\Thj{j}, \Phj{j}), \\
\psijb{j}{l}{m}(\vekrj{j}) &\equiv S_j(\vekr) \Njb{l}{j} \jj{l}(\kb \rj{j}) \Yj{l}{m}(\Thj{j}, \Phj{j}),
\end{align}
\end{subequations}
where
\begin{align}
S_j(\vekr) \equiv \begin{cases}
1 & \vekr \in V_{j} \\
0 & \text{otherwise}
\end{cases},
\end{align}
ensures that basis functions of different scatterers are orthogonal by construction. $\Nj{l}{j}$ and $\Njb{l}{j}$ are normalization constants, determined to satisfy \EqsrefM{Eqn:BasisIP}{Eqn:BackBasisIP}, and $\kj{j} \equiv \sqrt{\eps{j}} k_0$ and $\kb$ are the wave numbers of the $j$th scatterer and of the background medium, respectively. Finally, $\jj{l}(x)$ and $\Yj{l}{m}(\theta, \phi)$ are the spherical Bessel function of order $l$ and the spherical harmonic of degree $l$ and order $m$; These functions follow the conventions of~\cite{MultiScat}.

For representing the scalar Green's function, $\gB(\vekr,\vekr')$, we introduce the outgoing spherical wavefunctions
\begin{align} \label{Eqn:OutgoingWaveFunc}
\phib{l}{m}(\vekr) &\equiv \hj{l}(\kb r) \Yj{l}{m}(\theta, \phi),
\end{align}
where $\hj{l}(\kb \rj{j})$ is a spherical Hankel function of the first kind.

\section{Derivatives of Spherical Wavefunctions}\label{App:CarDerivSphWave}
We define the following functions
\begin{subequations}
\begin{align}
\psijNN{j}{l}{m}(\vekrj{j}) &\equiv\frac{ \psij{j}{l}{m}(\vekrj{j})}{\Nj{l}{j}}, \\ \psijbNN{j}{l}{m}(\vekrj{j}) &\equiv \frac{\psijb{j}{l}{m}(\vekrj{j})}{\Njb{l}{j}}
\end{align}
\end{subequations}
and let $\Omeg{l}{m}(\vekr)$ be any of these functions or the outgoing spherical wavefunctions
\begin{align}
\Omeg{l}{m}(\vekr) \in \Big \lbrace \psijNN{j}{l}{m}(\vekrj{j}), \, \psijbNN{j}{l}{m}(\vekrj{j}), \, \phib{l}{m}(\vekr) \Big \rbrace.
\end{align}
We also define the following differential operators~\cite{MultiScat}
\begin{subequations} \label{Eqn:DiffOperatorsDef}
\begin{align}
\mathcal{D}_1^{\pm} &\equiv -\frac{1}{k} \left(\partial_x \pm i \partial_y \right), \\
\mathcal{D}_1^{0} &\equiv -\frac{1}{k} \partial_z,
\end{align}
\end{subequations}
where $k$ is $k_j$ for $\psijNN{j}{l}{m}(\vekrj{j})$ and $\kb$ for $\psijbNN{j}{l}{m}(\vekrj{j})$ and $\phib{l}{m}(\vekr)$.

The action of these differential operators on $\Omeg{l}{m}(\vekr)$ can be expressed as~\cite{MultiScat}
\begin{subequations} \label{Eqn:DiffOperatorsAct}
\begin{align}
\mathcal{D}_1^{\pm} \Omeg{l}{m}(\vekr) &= \mp \left( \sqrt{\frac{(l\pm m +2) (l \pm m +1)}{(2l+1)(2l+3)}} \Omeg{l+1}{m\pm 1}(\vekr) \right. \nonumber \\
&\hspace{1cm}+ \left. \sqrt{\frac{(l \mp m) (l \mp m - 1)}{4l^2-1}} \Omeg{l-1}{m\pm 1}(\vekr) \right), \\
\mathcal{D}_1^{0} \Omeg{l}{m}(\vekr) &= \sqrt{\frac{(l+1)^2-m^2}{(2l+1)(2l+3)}} \Omeg{l+1}{m}(\vekr) \nonumber \\
&\hspace{1cm}- \sqrt{\frac{l^2-m^2}{4l^2-1}} \Omeg{l-1}{m}(\vekr).
\end{align}
\end{subequations}
Combining \EqsrefM{Eqn:DiffOperatorsDef}{Eqn:DiffOperatorsAct}, the detailed expressions for $\partial_{\alpha} \partial_{\alphap} \psijbNN{j}{\nu}{\mu}(\vekrj{j})$, for example, may be obtained.

\section{Separation Matrices} \label{App:SepMatrices}
The separation matrix, $\Smat{p}{t}{\nu}{\mu}(\vekb)$, introduced in \Eqref{Eqn:gBTwoCenter}, can be expressed as follows~\cite{MultiScat}
\begin{align}
\Smat{p}{t}{\nu}{\mu}(\vekb) = 4 \pi (-1)^{\nu+\mu+Q} &\sum_{q=0}^Q (-1)^{q} \phib{q_0+2q}{t-\mu}(\vekb) \nonumber \\
 &\times \Gaunt{p}{t}{\nu}{-\mu}{q_0+2q},
\end{align}
where we have employed the following definitions
\begin{subequations} \label{Eqn:SepMatVarious}
\begin{align}
Q &= \frac{p+\nu-q_0}{2}, \\
q_0 &= q_0(p,t; \nu, -\mu), \\
q_0(p, t; \nu, \mu) &= \begin{cases}
|p-\nu| & \text{if } |p-\nu| \geq |t+\mu| \\
|p+\mu| & \text{if } |p-\nu| < |t+\mu| \text{ and } \\
& p+\nu+|t+\mu| \text{ is even} \\ 
|p+\mu|+1 & \text{if } |p-\nu| < |t+\mu| \text{ and } \\
&  p+\nu+|t+\mu| \text{ is odd} 
\end{cases},\\
\Gaunt{p}{t}{\nu}{\mu}{q} &= (-1)^{t+\mu} \intSA \Yj{p}{t}(\theta,\phi) \Yj{\nu}{\mu}(\theta,\phi) \nonumber \\
&\hspace{2.5cm}\times \Yj{q}{-t-\mu}(\theta,\phi) \ddd \Omega, \label{Eqn:GauntDef}
\end{align}
with $\dd \Omega \equiv \sin(\theta) \dd \theta \dd \phi$, and $\Omega$ denoting $4\pi$ steradians. $\mathcal{G}$ is a Gaunt coefficient that is evaluated as follows
\begin{align}
\Gaunt{p}{t}{\nu}{\mu}{q} \equiv &(-1)^{t+\mu} \sqrt{\frac{(2p+1)(2\nu+1)(2q+1)}{4\pi}} \nonumber \\
&\times
\begin{pmatrix}
p & \nu & q \\
0 & 0 & 0
\end{pmatrix}
\begin{pmatrix}
p & \nu & q \\
t & \mu & -t-\mu
\end{pmatrix},
\end{align}
\end{subequations}
where the two last factors are so-called Wigner 3-$j$ symbols. $\Smath{p}{t}{\nu}{\mu}(\vekb)$, introduced in the expansion in \Eqref{Eqn:ExpanBasisrj}, can be expressed as follows~\cite{MultiScat}
\begin{align} \label{Eqn:ShatDef}
\Smath{p}{t}{\nu}{\mu}(\vekb) = 4 \pi (-1)^{\nu+\mu+Q} &\sum_{q=0}^Q (-1)^{q} \psijNN{j}{q_0+2q}{t-\mu}(\vekb) \nonumber \\
&\times \Gaunt{p}{t}{\nu}{-\mu}{q_0+2q}.
\end{align}

\section{Radial Integrals} \label{App:RadIntegrals}
We first consider the integral $\IA{j}{l}$ which is defined as
\begin{align} 
\IA{j}{l} \equiv \lim\limits_{\delta R' \rightarrow 0} \left( \int_{\delta R'}^{\infty} \hj{l}(\kb r') \jj{l}(\kj{j} r') r'^2 \ddd r' \right).
\end{align}
Expanding the integrand for small arguments, we find
\begin{align}
\lim\limits_{r' \rightarrow 0} \left( \hj{l}(\kb r') \jj{l}(\kj{j} r') r'^2 \right) = 0.
\end{align}
This implies that the integrand is bounded in the limit $r' \rightarrow 0$, and consequently we may evaluate the limit before expressing the integral
\begin{align}
\IA{j}{l} &= \int_{0}^{\infty} \hj{l}(\kb r') \jj{l}(\kj{j} r') r'^2 \ddd r' \nonumber \\
&= \frac{\pi}{2\sqrt{\kb \kj{j}}} \int_{0}^{\infty} H_{l+1/2}(\kb r') J_{l+1/2}(\kj{j} r') r' \ddd r'.
\end{align}
The remaining integral has been expressed in~\cite{Kellendonk}
\begin{align}
\IA{j}{l} &= -\frac{\ci}{\kb} \frac{1}{\kj{j}^2-\kb^2} \left( \frac{\kj{j}}{\kb} \right)^l.
\end{align}

Next, we express the finite part radial integral \cite{Schaums}
\begin{align}
I_l^{V_j} &\equiv \int_0^{\Rad{j}} \hj{l}(\kb r') \jj{l}(\kj{j} r') r'^2 \ddd r' \nonumber \\
&= \frac{M_l^j}{\Nj{j}{l} \Njb{j}{l}} + \ci \frac{\pi}{2} \sqrt{\frac{1}{\kj{j} \kb}} \frac{1}{\kj{j}^2-\kb^2} \nonumber \\
&\hspace{0.5cm}\times \bigg \lbrace \Big [\kj{j}\Rad{j} Y_{l+1/2}(\kb \Rad{j}) J_{l+3/2}(\kj{j} \Rad{j})  \nonumber \\
&\hspace{0.9cm} - \kb\Rad{j} J_{l+1/2}(\kj{j} \Rad{j}) Y_{l+3/2}(\kb \Rad{j})  \Big ] \nonumber \\
&\hspace{0.9cm} - \kj{j}^{l+1/2} \kb^{-l-1/2} \frac{2}{\pi}  \bigg \rbrace.
\end{align}

Finally, using the integrals expressed above, we have the radial integral needed to express the $B_{j,j}^{\alpha \alphap}$-integrals
\begin{align}
\IB{j}{l} \equiv \int_{\Rad{j}}^{\infty} \hj{l}(\kb r') \jj{l}(\kj{j} r') r'^2 \ddd r' = \IA{j}{l} - I_l^{V_j}.
\end{align}

\section{Expansion Coefficients of $\GB(\vekr,\vekr')$}\label{App:DipoleExpanCoefs}
To expand $\GB(\vekr,\vekr')$ on the background spherical wavefunctions, $\psijb{j}{l}{m}(\vekrj{j})$, cf. \Eqref{Eqn:FieldBExpan}, a slightly modified version of \Eqref{Eqn:BIntgB} for expressing the scalar Green's function, $\gB(\vekr,\vekr')$, is applied
\begin{align}
\gB(\vekr,\vekr') = \ci\kb \Dsum{\nu}{\mu} (-1)^{\mu} \psijbNN{j}{\nu}{\mu}(\vekrj{j}) \phib{\nu}{-\mu}(\vekrjp{j}).
\end{align}
We then express the elements of the background Green's tensor by the differential operators acting on $\phib{\nu}{-\mu}(\vekrjp{j})$
\begin{align}
\GB^{\alpha \alphap}(\vekr,\vekrd) &= \ci\kb \Dsum{\nu}{\mu} (-1)^{\mu} \psijb{j}{\nu}{\mu}(\vekrj{j}) / \left(\Njb{\nu}{j} \right) \nonumber \\
&\hspace{0.5cm}\times \bigg(\kd{\alpha}{\alphap} \phib{\nu}{-\mu}(\vekrjp{j}) + \frac{1}{\kb^2} \sum_{\gamma_{\alpha,\alphap}} g_{\gamma_{\alpha,\alphap}} \nonumber \\
&\hspace{0.9cm}\times \phib{\nu(\gamma_{\alpha,\alphap})}{\mu^*(\gamma_{\alpha,\alphap})}(\vekrjp{j}) \bigg),
\end{align}
where $\mu^*(\gamma_{\alpha,\alphap}) \equiv -\mu + \gamma''_{\alpha,\alphap}$. This relation is what we wanted: An expansion of the elements of the background Green's tensor on $\psijb{j}{\nu}{\mu}(\vekrj{j})$. The expansion coefficients are thus as given in \Eqref{Eqn:GBExpanCoefs}.

\section*{Acknowledgments}
This work was supported by the Villum Foundation via the VKR Centre of Excellence NATEC and the Danish Council for Independent Research (FTP 10-093651).




\end{document}